\documentclass[superscriptaddress,twocolumn,showpacs,pre,floatfix]{revtex4-1}

\usepackage{color}
\usepackage{tabularx}
\usepackage{epsfig}
\usepackage{amsmath}

\usepackage{graphicx}

\begin{document}

\title{Ising Spin Glasses in dimension two; universality and non-universality}

\author{P. H.~Lundow} 
\affiliation{Department of Mathematics and Mathematical Statistics,
  Ume{\aa} University, SE-901 87, Sweden}

\author{I. A.~Campbell}
\affiliation{Laboratoire Charles Coulomb, Universit\'e Montpellier II,
  34095 Montpellier, France}

\begin{abstract}
  Following numerous earlier studies, extensive simulations and
  analyses were made on the continuous interaction distribution
  Gaussian model and the discrete bimodal interaction distribution
  Ising Spin Glass (ISG) models in dimension two (P.H. Lundow and
  I.A. Campbell, Phys. Rev. E {\bf 93}, 022119 (2016)). Here we
  further analyse the bimodal and Gaussian data together with data on
  two other continuous interaction distribution 2D ISG models, the
  uniform and the Laplacian models, and three other discrete
  interaction distribution models, a diluted bimodal model, an
  "anti-diluted" model, and a more exotic symmetric Poisson
  model. Comparisons between the three continuous distribution models
  show that not only do they share the same exponent $\eta \equiv 0$
  but that to within the present numerical precision they share the
  same critical exponent $\nu$ also, and so lie in a single
  universality class. On the other hand the critical exponents of the
  four discrete distribution models are not the same as those of the
  continuous distributions, and differ from one discrete distribution
  model to another. Discrete distribution ISG models in dimension two
  have non-zero values of the critical exponent $\eta$; they do not
  lie in a single universality class.
\end{abstract}


\maketitle

\section{Introduction}\label{sec:I}
The canonical dimension $d = 2$ Edwards-Anderson (EA) model Ising spin
glasses (ISGs) on square lattices with either Gaussian or bimodal
($\pm J$) nearest neighbor interaction distributions have been the
subject of numerous studies over many years. Below we will refer in
particular to our own measurements on these two models
\cite{lundow:16}. There are analytic arguments that these two
archetype models (and by extension all 2D ISG models with other
distributions) have zero-temperature transitions
\cite{hartmann:01,ohzeki:09}.

After explaining the simulation and analysis techniques used, we first
present data on two other continuous distribution models : the uniform
and the Laplacian interaction distribution models, comparing with the
Gaussian model.  For the Gaussian model, where the interaction
distribution is continuous and the ground state for each individual
sample is unique, there is a general consensus concerning the
thermodynamic limit (ThL) critical exponents : $\eta \equiv 0$, $\nu =
3.52(2)$ \cite{rieger:97, hartmann:02, carter:02, hartmann:02a,
  houdayer:04, fernandez:16}.  We find that not only is the anomalous
dimension critical exponent $\eta \equiv 0$ for each of these three
models as it must be, but also that the correlation length exponent is
$\nu=3.52(5)$ for all three models to within the precision of the
present numerical data extrapolations. The data are thus compatible
with all 2D continuous interaction distribution models lying in a
single universality class.

For the 2D bimodal model the interaction distribution is discrete and
the ground state is highly degenerate. There are two limiting regimes,
with a size dependent crossover temperature $T^{*}(L)$ \cite{jorg:06},
a $T < T^{*}(L)$ ground state plus gap dominated regime and an
effectively continuous energy level regime $T > T^{*}(L)$. There have
been consistent estimates over decades from correlation function
measurements \cite{morgenstern:80,mcmillan:83}, Monte Carlo
renormalization-group measurements \cite{wang:88}, transfer matrix
calculations \cite{ozeki:90}, numerical simulations
\cite{houdayer:01,katzgraber:05,katzgraber:07,lundow:16}, and ground
state measurements \cite{poulter:05,hartmann:08} showing that the
anomalous dimension critical exponent $\eta \approx 0.20$ in both
regimes, indicating that the bimodal model is not in the same
universality class as the continuous distribution models. However, it
has also been claimed that the bimodal model in the $T > T^{*}(L)$
regime is in the same universality class as the Gaussian model,
because for the bimodal model : \lq\lq{}fits... lead to values of
$\eta$ that are very small, between 0 and 0.1, strongly suggestive of
$\eta = 0$\rq\rq{}~\cite{jorg:06}, and \lq\lq{}the data are not
sufficiently precise to provide a precise determination of $\eta$,
being consistent with a small value $\eta \leq 0.2$, including
$\eta=0$\rq\rq{}~\cite{parisen:10,parisen:11}. Recently the much more
definitive statement has been made : "we can safely summarize our
findings as $|\eta| < 0.02$."~\cite{fernandez:16}.

We discuss the Binder cumulant/correlation length ratio comparison
approach \cite{jorg:06a} in the 2D context, as applied to the
continuous interaction distribution models and to the bimodal model,
and then the Quotient approach used in Ref.~\cite{fernandez:16} as
applied to the bimodal model. From both approaches we deduce estimates
for the bimodal ISG exponents in the $T>T^{*}(L)$ regime which are
fully compatible with our previous conclusions Ref.~\cite{lundow:16}
including $\eta \approx 0.20$.

We then study three other discrete interaction distribution models : a
diluted bimodal ISG, an "anti-diluted" bimodal model and a
symmetric Poisson model. Using the approach of
Ref.~\cite{lundow:16} and the correlation length ratio/Binder cumulant
approach we conclude that each discrete interaction model has a
non-zero anomalous dimension exponent $\eta$ and lies in an individual
universality class.

\section{Simulations and analysis}\label{sec:II}
Simulations were carried out on square lattice Ising spin glasses
(ISGs) with near neighbor interactions, up to size $L=128$ and with $N
= 2^{13} = 8192$ independent samples at each size. Each of the 2D ISG
models orders only at zero temperature. As in Ref.~\cite{lundow:16}
where measurements were made on the square lattice ISG models with
Gaussian and bimodal interaction distributions, the samples were
equilibrated using the Houdayer method \cite{houdayer:01} with four
replicas; all the simulation techniques are identical to those already
described in detail in Ref.~\cite{lundow:16}.  As far as could be
judged by reading off the figures shown in Ref.~\cite{fernandez:16},
all the raw Gaussian and bimodal data in the \cite{fernandez:16} and
\cite{lundow:16} simulation sets are in full agreement with each other
to within the statistics.  For the present data analysis, in addition
to using $T$ as the temperature scaling variable, which is a standard
convention for models which order at zero temperature, we use
$\tau_{b} = 1/(1+\beta^2)$, where $\beta = 1/T$, as the scaling
variable (see Ref.~\cite{lundow:16}).  This variable is appropriate
for ISGs with $T_{c} = 0$ because of the symmetry between positive and
negative interactions in the distributions, and because $\tau_{b}$ has
the limits $\tau_{b} = 0$ at $T=0$, and $\tau_{b} = 1$ at infinite
temperature and so is well adapted to the Wegner scaling approach
\cite{wegner:72}. For consistency, when using this scaling variable we
scale not the bare second moment correlation length $\xi(\tau_{b},L)$
but the normalized correlation length $\xi(\tau_{b},L)/\beta$
following a general rule for ISGs in any dimension
\cite{campbell:06}. The normalized correlation length (like the
susceptibility $\chi(\tau_{b},L)$ and the normalized Binder cumulant
$g(\tau_{b},L) L^2$), tends to $1$ and not to $0$ at infinite
temperature; in consequence the behavior of $\xi(\tau_{b},L)/\beta$
over the entire temperature range can be expressed to good precision
using only a few finite Wegner correction terms.

For any distribution, for samples of size $L$ in the temperature range
where $L\gtrsim 7\xi(\tau_{b},L)$ all observables are practically
independent of $L$ and so can be considered to be in the Thermodynamic
limit (ThL) regime where observable values at finite $L$ are equal to
the infinite size limit values. This regime can be readily identified
by inspection of scaling plots.

In order to underline the validity of the analysis procedure which was
used for the bimodal and Gaussian ISG data in Ref.~\cite{lundow:16}
and which is again used below for the other ISG models, in Appendix I
we apply the same procedure to the Fully Frustrated (FF) Villain
model, a well understood 2D Ising model with a strongly degenerate
ground state which has a zero temperature ferromagnetic ordering point
and known critical behavior.

\section{The 2D continuous distribution ISG models}\label{sec:III}
The standard ISG Hamiltonian is $\mathcal{H}= -
\sum_{ij}J_{ij}S_{i}S_{j}$ with the near neighbor symmetric
distributions normalized to $\langle J_{ij}^2\rangle=1$.  The
normalized inverse temperature is $\beta = (\langle
J_{ij}^2\rangle/T^2)^{1/2}$.  The Ising spins are situated on simple
$L\times L$ grids with periodic boundary conditions. The spin overlap
parameter is defined as usual by
\begin{equation}
  q = \frac{1}{L^{d}}\sum_{i} S^{A}_{i}S^{B}_{i}
\end{equation}
where $A$ and $B$ indicate two copies of the same system and the sum
is over all sites.  The Laplacian interaction distribution is $P(J) =
\sqrt{2}\exp(-\sqrt{2}|J|)$, and the uniform interaction distribution
is $P(J)=1/(2\sqrt{3})$ for $-\sqrt{3} < J < \sqrt{3}$.  As in the
Gaussian distribution, these distributions are continuous in the
region around $J=0$; each sample has a unique ground state and an
anomalous dimension exponent $\eta \equiv 0$.

\begin{figure}
  \includegraphics[width=3.5in]{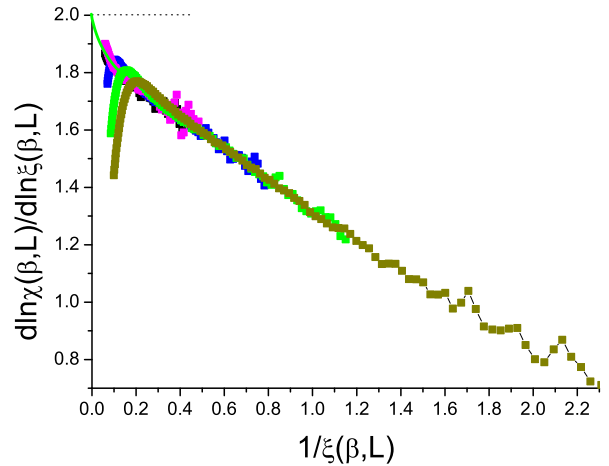} \caption{(Color on
    line) The logarithmic derivative of the SG susceptibility by the
    second moment correlation length
    $\partial\ln\chi(\beta,L)/\partial\ln\xi(\beta,L)$ against the
    inverse correlation length $1/\xi(\beta,L)$ for the Laplacian
    model. $L=128$, $96$, $64$, $48$, $32$ (left to right). Green
    continuous curve : extrapolation.} \protect\label{fig01}
\end{figure}

\begin{figure}
  \includegraphics[width=3.5in]{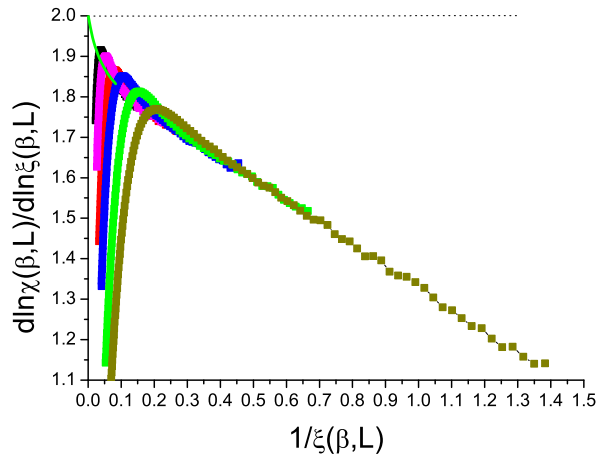} \caption{(Color on
    line) The derivative of the SG susceptibility by the second moment
    correlation length
    $\partial\ln\chi(\beta,L)/\partial\ln\xi(\beta,L)$ against the
    inverse correlation length $1/\xi(\beta,L)$ for the uniform
    model. $L=128$, $96$, $64$, $48$, $32$ (left to right). Green
    continuous curve : extrapolation. } \protect\label{fig02}
\end{figure}

We first show in Fig.~\ref{fig01} and Fig.~\ref{fig02} $y(\beta,L) =
\partial\ln\chi(\beta,L)/\partial\ln\xi(\beta,L)$ against $x(\beta,L)
= 1/\xi(\beta,L)$ for these two models; the data can be compared with
the data for the Gaussian model already shown in
Ref.~\cite{lundow:16}, Fig.~3. As must be the case for continuous
distributions, the ThL envelope for the derivative in each of these
models is consistent with an extrapolation to $y(\beta,L) = 2.0$ at
zero temperature $x(\beta,L)=0$, corresponding to the critical
exponent $\eta = 0$ in each model.

\begin{figure}
  \includegraphics[width=3.5in]{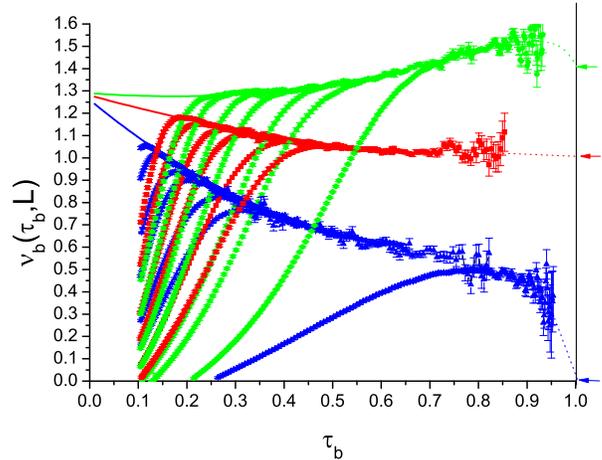} \caption{(Color on
    line) The logarithmic derivative of the normalized second moment
    correlation length
    $\partial\ln[\xi(\tau_{b})/\beta)]/\partial\ln\tau_{b}$ for the
    uniform (top sets, green circles), Gaussian (middle sets, red
    squares) and Laplacian (bottom sets, blue triangles), $L= 128$,
    $96$, $64$, $48$, $32$, $24$, $8$ (left to right in each
    case). Dashed curves : fits. Arrows : exact infinite temperature
    limits. } \protect\label{fig03}
\end{figure}

\begin{figure}
  \includegraphics[width=3.5in]{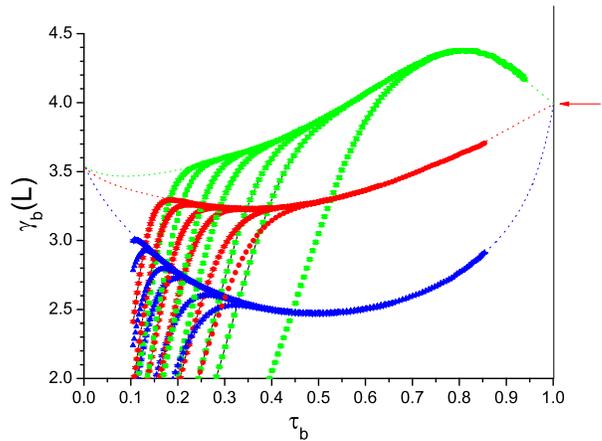} \caption{(Color on
    line) The logarithmic derivative of the spin glass susceptibility
    $\partial\ln\chi(\tau_{b})/\partial\ln\tau_{b}$ for the uniform
    (top sets, green circles), Gaussian (middle sets, red squares) and
    Laplacian (bottom sets, blue triangles), $L= 128,96,64,48,32,24,8$
    (left to right in each case). Dashed curves : fits. Arrow : exact
    infinite temperature limit for all distributions. }
  \protect\label{fig04}
\end{figure}

In Fig.~\ref{fig03} we show the effective correlation length exponents
$\nu_{b}(\beta,L) =
\partial\ln[\xi(\beta,L)/\beta]/\partial\ln\tau(\beta)$ as functions
of $\tau_{b}$ together for all sizes $L$ and for all three continuous
distribution models.  In Fig.~\ref{fig04} we show the effective
susceptibility exponents $\gamma_{b}(\beta,L) =
\partial\ln\chi(\beta,L)/\partial\ln\tau(\beta)$ again for all $L$ and
for all three models.  We have carried out extrapolations using just
the same polynomial fit procedure as explained in detail in
\cite{lundow:16} and in the Appendix. The extrapolated zero
temperature critical exponent estimates are $\nu_{b} = 1.27(2)$ and
$\gamma_{b} = 3.52(5)$ for all three models. For all models
(continuous and discrete interaction distributions) these critical
exponents are related to the correlation length $\nu$ and anomalous
dimension $\eta$ critical exponents in the traditional $T$ scaling
convention by $\nu_{b} = (\nu - 1)/2$ and $\gamma_{b} =
\nu(2-\eta)/2$. The exact infinite temperature limits are $\nu_{b} =
2- K/3$ where $K$ is the kurtosis of the interaction distribution, and
$\gamma_{b} = 4$ \cite{lundow:16}.


Thus all the critical exponent estimates for these three
non-degenerate ground state models are compatible with $\eta = 0$ and
$\nu = 3.52(2)$. We conclude that all two-dimensional non-degenerate
ground-state ISG models lie in a single universality class; not only
is $\eta = 0$ which must be true for this class of models, but also
all critical $\nu$ values appear to be identical within the
statistical and extrapolation errors. The strength and sign of the
corrections to scaling are, however, quite different for the different
models. Again, with the $\tau_{b}$ scaling convention, the correlation
lengths with the leading Wegner scaling corrections assuming a leading
correction exponent $\theta =1$ are
\begin{equation}
  \xi(\tau_{b}) = (0.69/\beta)\tau_{b}^{-1.28}[1 + 0.49\tau_{b} + \cdots]
  \label{xx}
\end{equation}
for the Gaussian model,
\begin{equation}
  \xi(\tau_{b}) = (1.13/\beta)\tau_{b}^{-1.28}[1 - 0.04\tau_{b} + \cdots]
  \label{xy}
\end{equation}
for the uniform model, and
\begin{equation}
  \xi(\tau_{b}) = (0.25/\beta)\tau_{b}^{-1.28}[1 + 2.5\tau_{b} + \cdots]
  \label{xz}
\end{equation}
for the Laplacian model.

It can be noted that these data provide a validation of the
extrapolation procedure outlined in \cite{lundow:16} and in the
Appendix. Although the corrections are very different in the three
models, the extrapolations to criticality lead to consistent exponent
values. {\it A priori} this implies that for other models where the
same extrapolation procedure leads to other critical exponent
estimates, these different values can be considered to be reliable.

\section{Correlation length ratio and Binder cumulant scaling}\label{sec:IV}
Universality in ISGs has been tested through comparing plots of the
Binder parameter $g(\beta,L)$ against the second moment correlation
length ratio $\xi(\beta,L)/L$ for different models, interpreted using
finite size scaling arguments (see for instance Ref.~\cite{jorg:06a}).

We will consider this type of scaling plot in the 2D context. In this
section we will use $U_{4}(\beta,L) = 3-2g(\beta,L)$ rather than
$g(\beta,L)$ to facilitate comparisons with Ref.~\cite{fernandez:16}.

Quite generally the 2D correlation function (either a spin-spin
correlation function for ferromagnets or a spin glass correlation
function for ISGs) at distance $r$ takes the asymptotic form
\begin{equation}
  G(\beta,r) \sim r^{-\eta}\exp[-r/\Xi(\beta)]
\end{equation}
with possible small $r$ finite size deviations, where $\Xi(\beta)$ is
the exponential or ``true'' correlation length (not the second moment
correlation length \cite{butera:04}). Dimensionless observables
$Q(\beta,L)$ such as $U_{4}(\beta,L)$ or $\xi(\beta,L)/L$ will each be
given by a general toroidal integral $Q(\beta,L) = \int^{L}
F_{Q}(r)G(\beta,r)r^{2} \mathrm{d}r$ where $F_{Q}(r)$ is the appropriate
function for the variable, or a ratio of integrals.

For any model with $\eta = 0$ so $G(\beta,r) \sim
\exp[-r/\Xi(\beta)]$, at given $\beta$ and $L$ the integrals are
entirely determined by $\Xi(\beta)$ and $L$ so whatever the
temperature variations of $\Xi(\beta)$ for a particular model, plots
of one dimensionless observable $Q_{a}(\beta,L)$ against another
dimensionless observable $Q_{b}(\beta,L)$ will be universal,
independent of the model and of $L$, in agreement with the general ISG
scaling rule \cite{jorg:06a}.  As the 2D models have $T_{c}=0$ the
universal curve for $\eta=0$ models will extend up to the critical
zero temperature limit $[U_{4}(0,L)=1, \xi(0,L)/L = \infty]$ for all
$L$.

\begin{figure}
  \includegraphics[width=3.5in]{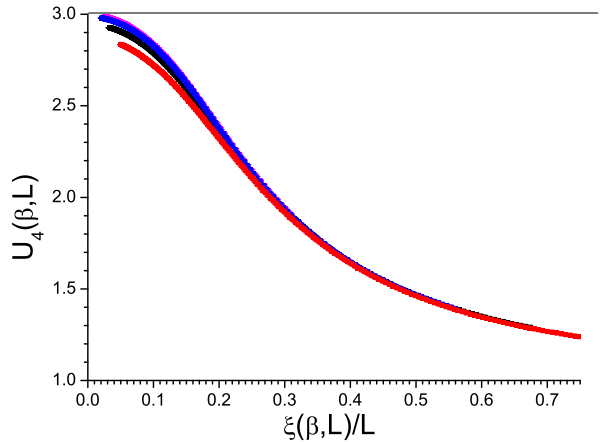} \caption{(Color on
    line) Plot of the Binder cumulant $U_{4}(\beta,L)$ against
    $\xi(\beta,L)/L$ for the 2D Laplacian model from $\xi(\beta,L)/L =
    0$ to $\xi(\beta,L)/L = 0.7$.  $L= 48$, $12$, $6$, $4$ (top to
    bottom). For all $L$ the curves will extend to $U_{4}(\beta,L)=1,
    \xi(\beta,L)/L = \infty$ at $T=0$. } \protect\label{fig05}
\end{figure}

\begin{figure}
  \includegraphics[width=3.5in]{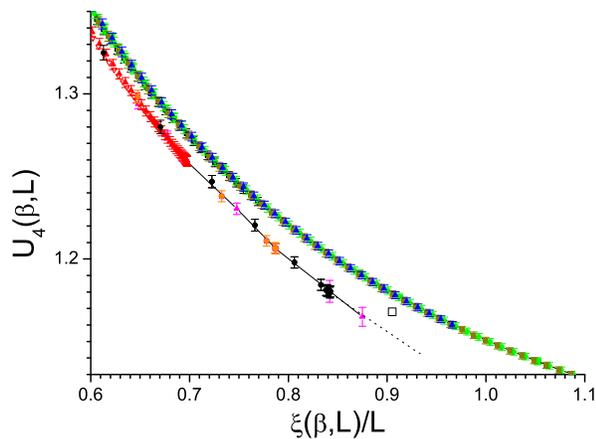}
  \caption{(Color on line) Plot of the Binder cumulant
    $U_{4}(\beta,L)$ against $\xi(\beta,L)/L$ from $\xi(\beta,L)/L =
    0.6$ to $\xi(\beta,L)/L = 1.1$.  In the top curve, the 2D $L=12$
    Gaussian model (green circles), the 2D $L=12$ uniform model (brown
    squares) and the 2D $L=12$ Laplacian model (blue triangles), all
    overlapping.  Lower set : the bimodal model, $L=12$ red inverted
    triangles, $L=16$ red triangles, $L=32$ orange diamonds, $L=48$
    black circles, $L=64$ pink triangles. Critical point 2D Ising
    ferromagnet : open square. } \protect\label{fig06}
\end{figure}

The measurements on the $\eta = 0$ ISG models show that for small to
moderate $L$ and $\xi(\beta)/L< 0.3$, the $U_{4}(\beta,L)$ against
$\xi(\beta,L)/L$ curves are not quite independent of $L$,
Fig.~\ref{fig05}. The small $L$ deviations can be ascribed to the
presence of pre-asymptotic corrections to $G(r)$. However, for
$\xi(\beta)/L > 0.3$, the $U_{4}(\beta,L)$ against $\xi(\beta,L)/L$
scaling curves for the Gaussian, uniform and Laplacian $\eta=0$ ISG
models become identical and independent of $L$ to within the
statistics, Fig.~\ref{fig06}. Only at very small sizes, $L \approx 4$,
are there still weak finite size deviations, which were seen also in
Ref.~\cite{fernandez:16} for the Gaussian model. The present data show
$L=4$ deviations for the uniform model which are very similar in
strength to the Gaussian deviations; the Laplacian model deviations
are rather weaker.

In any model where $\eta$ is not zero, at criticality
$\Xi(\beta_{c})=\infty$ and the critical observables will be given by
integrals with the asymptotic correlation function $G(\beta_{c},r)
\sim r^{-\eta}$. (As this function diverges at $r=0$, it must take up
an appropriate functional form such as $G(\beta_{c},r)=1/(1+r^{\eta})$
for small $r$, leading to small $L$ corrections). The explicit
infinite size critical toroidal integrals for the 2D Ising ferromagnet
with $\eta \equiv 1/4$ were calculated by Salas and Sokal
\cite{salas:00}, and gave $\xi(\beta_{c},L)/L = 0.9050488292(4)$ and
$U_{4}(\beta_{c}) = 1.16792\ldots$. For the 2D Fully Frustrated model
with $\eta = 1/2$, from simulations there is a critical zero
temperature end-point at $\xi(0,L)/L = 0.49(1)$, $U_{4}(0,L)=1.615(5)$
(\cite{katzgraber:08} and see Appendix I), with weak finite size
effects.  Numerical toroidal integrations for critical points could in
principle be carried out for other $\eta$ values. In 2D strip geometry
at criticality $\xi(\beta_{c},L)/L = 1/(\pi\eta)$ \cite{cardy:84}. The
Ising, FF and $\eta=0$ values in square geometry correspond
approximately to $\xi(\beta_{c},L)/L = 1/(4.4\eta)$ , and we can take
this as a rough calibration for the estimation of the ISG $\eta$
values from end-point $\xi(0,L)/L$ estimates.(Unfortunately all other
partially frustrated $S=1/2$ 2D Ising models have finite ordering
temperatures and $\eta = 1/4$ like the Ising model \cite{wu:03} so can
give no further critical point information).


For non-zero $\eta$ ISG
models with $T_{c}=0$ one can expect $[U_{4}(0,L), \xi(0,L)/L]$
end-point limits for each $L$, with a critical zero temperature
end-point limit for infinite $L$ whose location will be determined
uniquely by $\eta$.

In Fig.~\ref{fig06}, $[U_{4}(0,L), \xi(0,L)/L]$ scaling plots are
compared.  In addition to a part of the $\eta=0$ ISG universal scaling
curve we show the 2D Ising ferromagnet $T_{c}$ critical point, and
scaling data for the 2D bimodal ISG.  The Ising ferromagnet
$\eta=0.25$ critical point happens to lie rather close to the
universal $\eta=0$ curve. For the bimodal ISG model, data for each $L$
can be seen to leave a common $\Xi(\beta)$ dominated regime curve
(which is similar to but distinct from the $\eta=0$ universal curve)
before smoothly attaining a weakly $L$ dependent end-point,
corresponding to the $T < T^{*}(L)$ ground state regime. The
observation that for each $L$ this behavior is smooth and regular as
the temperature tends to zero, with a final bunching up of data points
when the ground state regime is reached, shows that the effective
$\eta$ in the $T>T^{*}(L)$ regime and in the (weakly $L$-dependent)
$T<T^{*}(L)$ ground state regime are essentially the same. In other
words the state degeneracy and hence $\eta$ depends only mildly on
temperature, right through the $T^{*}(L)$ crossover. The series of
end-points for increasing $L$ will terminate at an infinite $L$
bimodal model end-point (see Ref.~\cite{katzgraber:07}) which is close
to but beyond the ferromagnetic Ising critical point, so consistent
with a bimodal ISG $\eta$ which is lower than but close to $\eta =
0.25$. By inspection, the bimodal ISG data are totally incompatible
with a critical exponent $\eta =0$.  The position of the infinite $L$
bimodal ISG end-point will be estimated below together with the
positions for three other discrete interaction distribution 2D ISG
models, Fig.~\ref{fig16}.

\section{The 2D bimodal ISG : the Quotient approach}\label{sec:V}
In Ref.~\cite{fernandez:16} raw 2D Gaussian and bimodal ISG simulation
data broadly equivalent to 
Ref.~\cite{lundow:16} were generated; these were analysed using a
Quotient approach, with the normalized second moment correlation
length $x=\xi(T,L)/L$ as the scaling variable. It should be noted that
the Quotients in Ref.~\cite{fernandez:16} are at constant $x$ not
Quotients at constant $T$ as in for instance
Ref.~\cite{ballesteros:00}. Unfortunately no derivations are given in
Ref.~\cite{fernandez:16} for any of the important Quotient limit
expressions which are cited.  Here we provide simple derivations for
the Quotient limits and we discuss plots made up of data formatted
following the Quotient approach.


Assume the basic $T_{c}=0$ scaling expressions $\xi(T)\sim T^{-\nu}$
and $\chi(T) \sim T^{-(2-\eta)\nu}$, valid near the large $L$, $T \to
0$ critical limit.  At size $L$ and temperature $T$, $x(T,L) =
\xi(T,L)/L \sim T(x,L)^{-\nu}/L$.

Then for size $2L$ at the same $x$ and at temperature $T''(x,2L)$,
\begin{eqnarray}
  x(T'',2L) &=& \frac{\xi(T'',2L)}{2L} = \frac{T''(x,2L)^{-\nu}}{2L}
  \nonumber\\ &=&\frac{T(x,L)^{-\nu}}{L}
\end{eqnarray}
with $x(T,L)=x(T'',2L)$ ; so $ T(x,L)^{-\nu}/T''(x,2L)^{-\nu} = 2$
i.e. the Quotient $Q_{T}$ as defined in Ref.~\cite{fernandez:16} tends
to
\begin{equation}
  Q_{T} =  \frac{T''(x,2L)}{T(x,L)} = 2^{-1/\nu}
\end{equation}
in the large $L$ limit. This expression is identical to the limit
relation cited in Ref.~\cite{fernandez:16} Eqn.~(7), implying that the
limit derivation procedure followed was the same as the present
one. Using this expression, the Gaussian $Q_{T}(0)=0.82$ large $L$
intercept reported in Ref.~\cite{fernandez:16} is consistent with the
accepted literature value $\nu = 3.55(2)$
\cite{rieger:97,hartmann:02,carter:02,hartmann:02a,houdayer:04} for
the Gaussian ISG critical exponent.

Then
\begin{equation}
  \langle q^2\rangle(T,L) = \frac{\chi(T,L)}{L^2} =  \frac{T^{-(2-\eta)\nu}}{L^2}
\end{equation}
With $x = \xi(L,T)/L$, from above $T(x,L)^{-\nu} = \xi(T,L) =
Lx(T,L)$, so
\begin{eqnarray}
  \langle q^2\rangle(x,L) &=& \frac{T(x,L)^{-\nu (2-\eta)}}{L^2} =
  \frac{(xL)^{(2-\eta)}}{L^2} \nonumber\\ &=& L(x,T)^{-\eta}x^{2-\eta}
\end{eqnarray}
i.e. the Quotient $Q_{q^2} = \langle q^2\rangle(x,2L)/\langle
q^2\rangle(x,L) = 2^{-\eta}$ in the large $L$ limit. This is identical
to the expression cited in Ref.~\cite{fernandez:16}, Eqn.~(D3).

\begin{figure}
  \includegraphics[width=3.5in]{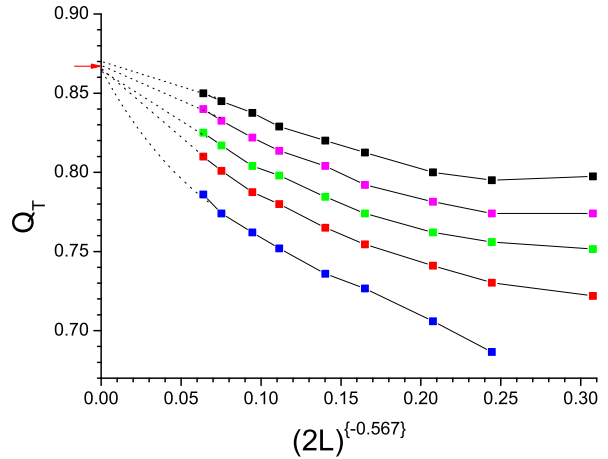}
  \caption{(Color on line) The 2D bimodal Quotient $Q_{T}(x,L)$ for $x
    = \xi(L)/L$ values $x = 0.1$, $0.2$, $0.3$, $0.4$, $0.5$ (bottom
    to top). The horizontal axis is $(2L)^{-0.567}$ as in
    Ref.~\cite{fernandez:16} (In this reference the axis is stated to
    be $(L)^{-0.567}$ which is incorrect).} \protect\label{fig07}
\end{figure}

\begin{figure}
  \includegraphics[width=3.5in]{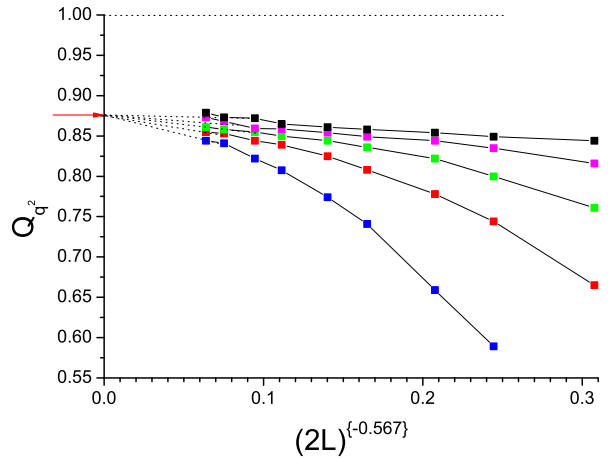} \caption{(Color on
    line) The 2D bimodal Quotient $Q_{q^2}(x,L)$ for $x = \xi(L)/L$
    values $x =0.1$, $0.2$, $0.3$, $0.4$, $0.5$ (bottom to top). The
    horizontal axis is $(2L)^{-0.567}$ as in Ref.~\cite{fernandez:16}
    (In this reference the axis is stated to be $(L)^{-0.567}$ which
    is incorrect). } \protect\label{fig08}
\end{figure}

We can inspect Figs.~\ref{fig07} and \ref{fig08} for the bimodal ISG
Quotients with points compiled from the present numerical data; the
figures are presented in just the same form as
Ref.~\cite{fernandez:16} Fig.~7 upper and middle. As far as can be
judged by reading off the plots in Ref.~\cite{fernandez:16}, point by
point agreement between the present Quotients and those of
Ref.~\cite{fernandez:16} is excellent (as could be expected as the raw
data should be essentially the same). The natural extrapolations
indicated in the present figures lead to bimodal ISG critical
infinite-$L$ Quotient intercept estimates $Q_{T}(0)= 0.865(10)$ and
$Q_{q^2}(0) = 0.87(1)$. (No equivalent extrapolations of the bimodal
Quotient data were made in Ref.~\cite{fernandez:16}, but if these had
been made the intercept estimates would have been very similar to the
present values). From the limit expressions above, these intercepts
correspond to bimodal critical exponent estimates $\nu = 4.8(3)$ and
$\eta = 0.20(2)$, estimates which are fully consistent with the
bimodal exponents estimated through a completely independent analysis
procedure in Ref.~\cite{lundow:16}. In particular the value obtained
for $\eta$ is clearly non-zero.

Finally, in Ref.~\cite{fernandez:16} section VI and Appendix C an
observable $g(x,T)$ is defined by $g(x,T) = \langle q^2\rangle
(x=0.4,T)/\langle q^2\rangle (x,T)$ averaged over $T$. (The factor
$[\hat{u}_{h}(T)]^2$ depends only on $T$ and so cancels out in the
ratio in Ref.~\cite{fernandez:16} Fig.~3).  Note that the $\langle
q^2\rangle (x=0.4,T)$ and $\langle q^2\rangle (x,T)$ in the definition
of $g(x,T)$ correspond to the same $T$ but at quite different $L$, say
$L(x,T)$ and $L''(0.4,T)$.

From the Quotient discussion for $Q_{q^2}$ above and assuming some
fixed $T$ : $\langle q^2\rangle (x,T) = L(x,T)^{-\eta}x^{2-\eta}$ and
from the $Q_{T}$ discussion $L(x,T) = T(x,L)^{-\nu}/x(T,L)$.

So :
\begin{eqnarray}
  \langle q^2\rangle (x=0.4,T) &=&
          [\overbrace{T(0.4,L'')^{-\nu}/0.4}^{T(x,L'')^{-\nu}/x(T,L'')}]^{-\eta} 0.4^{2-\eta}
          \nonumber\\ &=& [T(0.4,L'')^{-\nu}]^{-\eta} 0.4^{\eta}
          0.4^{2-\eta} \nonumber\\ &=& [T(0.4,L'')^{-\nu}]^{-\eta}
          0.4^2
\end{eqnarray}
and
\begin{eqnarray}
  \langle q^2\rangle (x,T) &=& [T(x,L)^{-\nu}/x(T,L)]^{-\eta}x^{2-\eta}
  \nonumber\\ &=& [T(x,L)^{-\nu}]^{-\eta} x^{\eta} x^{2-\eta} 
  \nonumber\\ &=&
  [T(x,L)^{-\nu}]^{-\eta} x^2
\end{eqnarray}

As $T(0.4,L'') = T(x,L)$,
\begin{equation}
  g(x,T) = \frac{\langle q^2\rangle (0.4,T)}{\langle q^2\rangle (x,T)} =
  0.4^{2} x^{-2} = \frac{0.16}{x^{2}}
\end{equation}
at small $x$ whatever $\eta$.  The log-log $g(x,T)$ against $x$ data
plot shown in Ref.~\cite{fernandez:16} Fig.~3 is entirely consistent
with this simple rule (including the pre-factor $0.16$) from $x=0.1$
to about $x=0.5$ for both the Gaussian and the bimodal models.

The relation $g(x) \sim 1/x^{2-\eta}$ cited (with no derivation) in
Ref.~\cite{fernandez:16} is in disagreement with the present
derivation, and with the observed data shown in
Ref.~\cite{fernandez:16}. The conclusion in Ref.~\cite{fernandez:16}
that $|\eta| < 0.02$ for the 2D bimodal ISG model, drawn principally
from $g(x,T)$ analyses, seems to have been based on an incorrect
expression and so is invalid.

To summarize, when the Quotient analyses presented in
Ref.~\cite{fernandez:16} with the limit derivations given above are
applied to the bimodal simulation data, estimates for the critical
exponents in the bimodal ISG model obtained by extrapolations of
$Q(T)$ and $Q(q^2)$ to large $L$ are consistent with those obtained
following the analysis procedure used in Ref.~\cite{lundow:16}. Both
bimodal exponents are quite different from the values for the
continuous distribution models. The $g(x,T)$ data analysis provides no
information on the critical exponents.

\section{Discrete interaction distribution ISGs}\label{sec:VI}
\begin{figure}
  \includegraphics[width=3.5in]{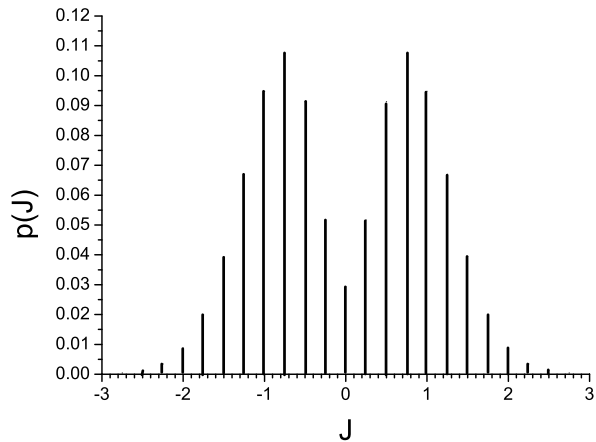}
  \caption{(Color on line) The interaction distribution for the
    symmetric Poisson ISG model.} \protect\label{fig09}
\end{figure}

Having studied the standard 2D bimodal model in \cite{lundow:16}, we
have now made equivalent measurements on three different degenerate
ground state models : a diluted bimodal model with a fraction
$p=0.125$ of the interactions set randomly to zero (a diluted bimodal
model was already studied in Refs.~\cite{lukic:06,hartmann:08}), an
"anti-diluted" bimodal model where a fraction $p=0.2$ of the
interactions are set randomly to strength $\pm 2J$ and the remaining
fraction to $\pm J$. Also we test a more complex symmetric Poisson
model with an interaction distribution shown in Fig.~\ref{fig09}; this
model has probability $\lambda^{|k|}\exp(-\lambda)/2$ for strength
$(k/4)J$, when $k\neq 0$, and probability $\exp(-\lambda)$ when $k=0$,
with $\lambda=(\sqrt{65}-1)/2$.

\begin{figure}
  \includegraphics[width=3.5in]{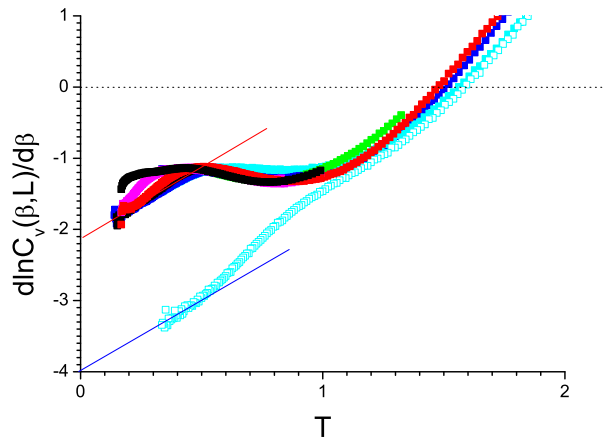}
  \caption{(Color on line) Diluted bimodal 2D ISG. Logarithmic
    derivative of the specific heat
    $\partial\ln C_{v}(\beta,L)/\partial\beta$ against $T$. Full
    points : $L= 32$, $24$, $16$, $12$, $6$, $4$ (green, black, pink,
    red, blue, cyan; top to bottom on the right). Open points :
    bimodal 2D ISG $L=4$ for comparison. Red line : $y(x) = -2.1+2x$,
    blue (lower) line $y(x) = -4+2x$.} \protect\label{fig10}
\end{figure}

\begin{figure}
  \includegraphics[width=3.5in]{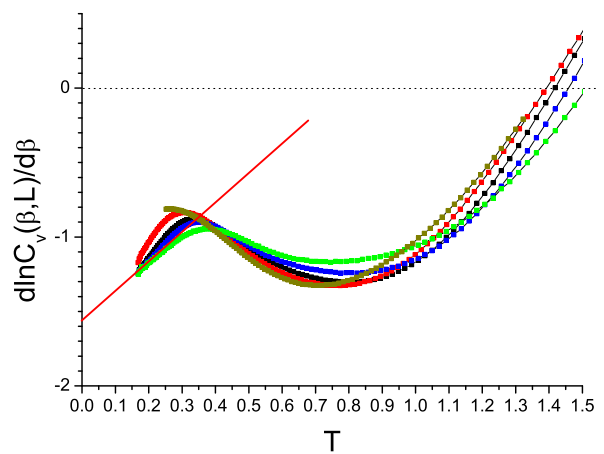}
  \caption{(Color on line) "Anti-diluted" bimodal 2D ISG. Logarithmic
    derivative of the specific heat
    $\partial\ln C_{v}(\beta,L)/\partial\beta$ against $T$. Full
    points : $L= 24$, $12$, $8$, $6$, $4$ (brown, red, black, blue,
    green; top to bottom on the right). Red line : $y(x) = -1.55+2x$.}
  \protect\label{fig11}
\end{figure}

\begin{figure}
  \includegraphics[width=3.5in]{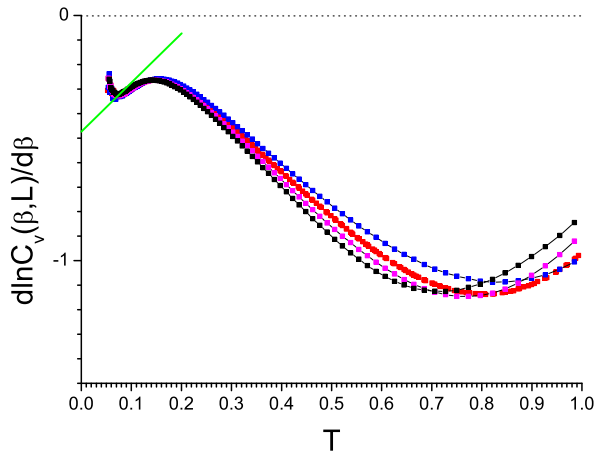}
  \caption{(Color on line) Symmetric Poisson 2D ISG. Logarithmic
    derivative of the specific heat
    $\partial\ln C_{v}(\beta,L)/\partial\beta$ against $T$. Full
    points : $L= 24$, $12$, $8$, $6$ (black, pink, red, blue; top to
    bottom on the right). Green line : $y(x) = -0.45+2x$.}
  \protect\label{fig12}
\end{figure}

These models have discrete interaction distributions and so can be
expected to have degenerate ground states; we do not, however, know the
values of the ground state degeneracy.  Logarithmic derivatives of the
specific heat data are shown in Figs.~\ref{fig10}, \ref{fig11} and
\ref{fig12} in the same format, $\partial\ln C_{v}/\partial\beta$
against $T$, as that of the bimodal ISG model in \cite{lundow:16}
Fig.~4 and of the FF model, Fig.~\ref{fig26} below. Again the discrete
distribution data indicate crossovers for all models, with a $T <
T^{*}(L)$ ground state plus gap regime specific heat of the form
$C_{v} \sim \beta^B\exp(-A\beta)$ having $B \approx 2$. The effective
gap parameter $A \approx 2.1$ for the diluted bimodal model, $A
\approx 1.5$ for the anti-diluted bimodal model, and $A\approx 0.5$
for the symmetric Poisson model, so significantly smaller than the gap
$A=4$ of both the FF and pure bimodal models. Ref.~\cite{lukic:06}
showed data on a perturbed FF model which were also interpreted as
having a gap $A$ weaker than $4$.  We do not dispose of large $L$ data
to low enough $T$ to be able to establish the limiting infinite size
$T > T^{*}(L)$ ThL form of $C_{v}(T)$ for these models.

\begin{figure}
  \includegraphics[width=3.5in]{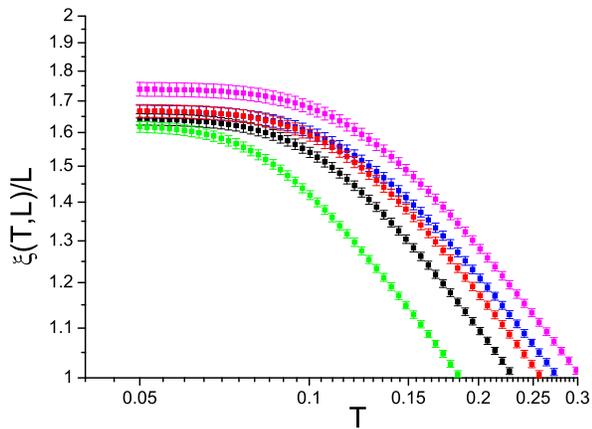}
  \caption{(Color on line) Symmetric Poisson 2D ISG. The normalized
    correlation length $\xi(T,L)/L$ against $T$ at low
    temperatures. $L= 4$, $6$, $8$, $12$, $24$ (top to bottom).}
  \protect\label{fig13}
\end{figure}

\begin{figure}
  \includegraphics[width=3.5in]{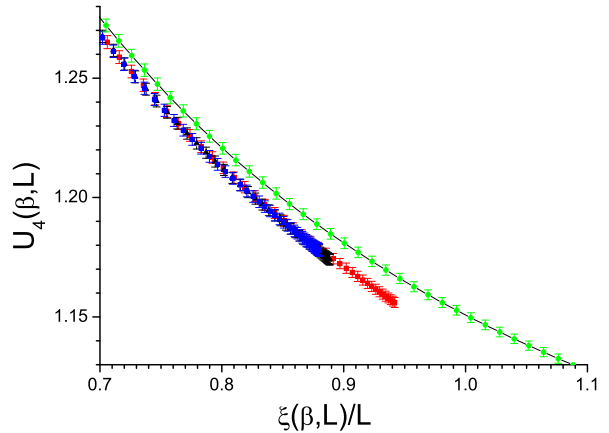}
  \caption{(Color on line) Diluted bimodal 2D ISG. The low temperature
    Binder cumulant $U_{4}(\beta,L)$ against the normalized
    correlation length $\xi(\beta,L)/L$. ($L=24$ red squares, $L=12$
    black triangles, $L=8$ blue diamonds).  For each $L$ the data
    points terminate at a zero temperature end-point. For comparison,
    the universal continuous distribution curve is represented by
    $L=12$ Gaussian ISG data (upper set, green circles) which extend
    to infinity.}  \protect\label{fig14}
\end{figure}

\begin{figure}
  \includegraphics[width=3.5in]{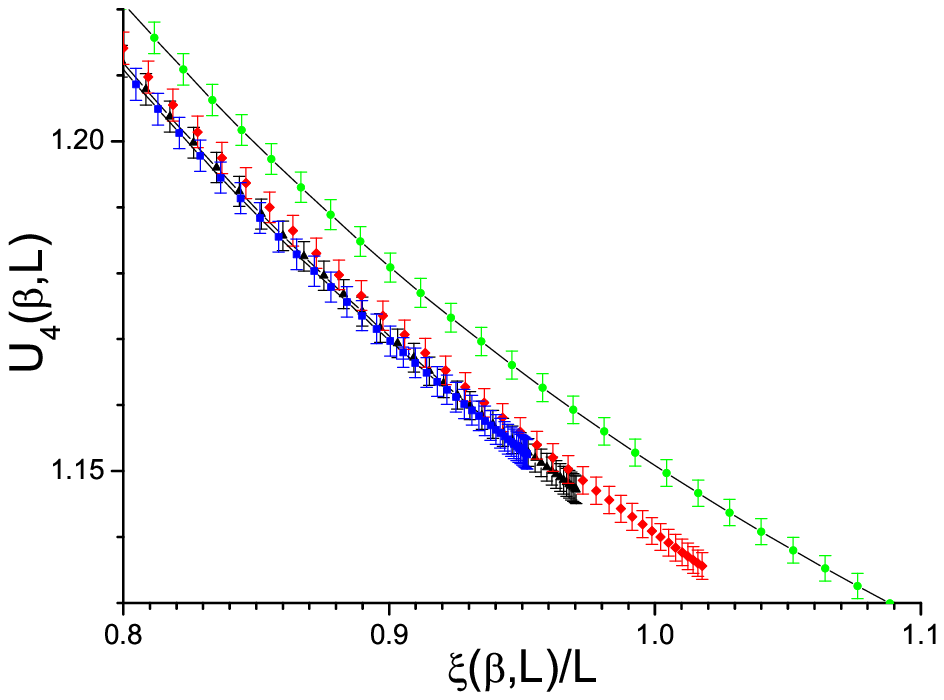}
  \caption{(Color on line) Anti-diluted bimodal 2D ISG. The low
    temperature Binder cumulant $U_{4}(\beta,L)$ against the
    normalized correlation length $\xi(\beta,L)/L$. ($L=12$ red
    diamonds, $L=8$ black triangles, $L=6$ blue squares). For each $L$
    the data points terminate at a zero temperature end-point. For
    comparison, the universal continuous distribution curve is
    represented by $L=12$ Gaussian ISG data (upper set, green circles)
    which extend to infinity. } \protect\label{fig15}
\end{figure}

\begin{figure}
  \includegraphics[width=3.5in]{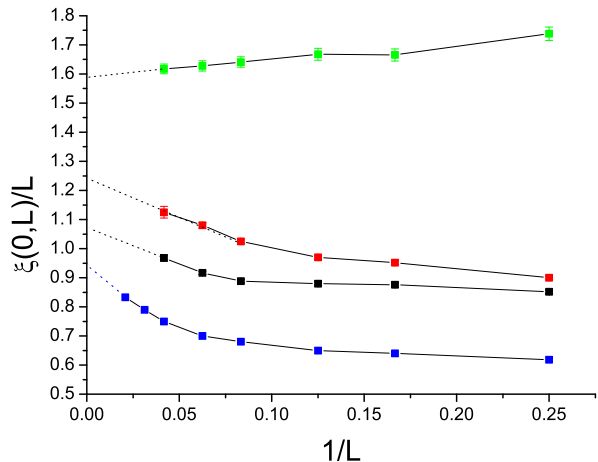}
  \caption{(Color on line) The size dependent zero temperature
    end-point values of $\xi(\beta,L)/L$ for the four discrete
    interaction models : the symmetric Poisson model, the anti-diluted
    bimodal model, the diluted bimodal model, and the bimodal model,
    from top to bottom.} \protect\label{fig16}
\end{figure}

In each of the discrete interaction models, the normalized correlation
length saturates at an end-point value at low temperature for all
$L$. As an example the data for the symmetric Poisson model are shown
in Fig.~\ref{fig13}. Binder cumulant $U_{4}(\beta,L)$ against
normalized correlation length $\xi(\beta,L)/L$ plots are shown in
Figs.~\ref{fig14} and \ref{fig15} for the diluted bimodal and
anti-diluted bimodal models.  As for the bimodal model the data points
lie on a curve distinct from the continuous distribution universal
curve and tend to end-points for each $L$ at zero temperature,
behavior characteristic of a non-zero exponent $\eta$. The end-point
values of $\xi(0,L)/L$ for all four discrete interaction models are
shown plotted against $1/L$ in Fig.~\ref{fig16}. The infinite $L$
end-point values estimated by extrapolation are distinct, indicating
that the $\eta$ values are distinct so the discrete interaction models
are all in different universality classes.

From the approximate calibration of the $[\xi(L)/L]_{T=0}$ infinite
$L$ end-point values in terms of $\eta$ above, we can give estimates
$\eta \approx 0.24$, $0.21$, $0.18$, $0.14$ respectively for the
bimodal, diluted, anti-diluted and symmetric Poisson models. We can
remark that the end point values lie close to but beyond the 2D Ising
ferromagnet critical value, implying the ISG $\eta$ values are all
near to but somewhat below $0.25$. The $\eta$ values are roughly
consistent with the $\eta$ estimates from a diferent approach given
below.


\begin{figure}
  \includegraphics[width=3.5in]{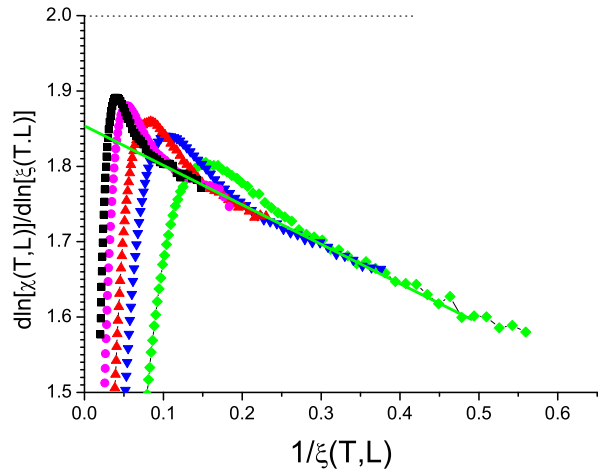}
  \caption{(Color on line) Diluted bimodal 2D ISG. Logarithmic
    derivative $\partial\ln\chi(\beta,L)/\partial\ln\xi(\beta,L)$
    against $1/\xi(\beta,L)$ for $L= 128$, $96$, $64$, $48$, $32$
    (black, pink, red, blue, green) left to right. The continuous
    (green) curve is an extrapolated fit.}  \protect\label{fig17}
\end{figure}

\begin{figure}
  \includegraphics[width=3.5in]{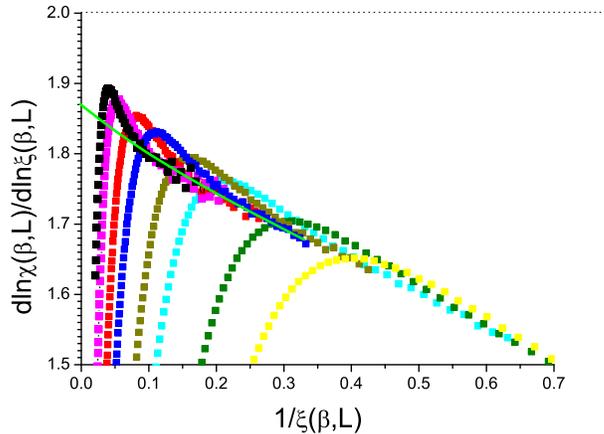}
  \caption{(Color on line) Anti-diluted bimodal 2D ISG. Logarithmic
    derivative $\partial\ln\chi(\beta,L)/\partial\ln\xi(\beta,L)$
    against $1/\xi(\beta,L)$ for $L= 128$, $96$, $48$, $32$, $24$,
    $16$, $12$ (left to right). The continuous (green) curve is an
    extrapolated fit.}  \protect\label{fig18}
\end{figure}

\begin{figure}
  \includegraphics[width=3.5in]{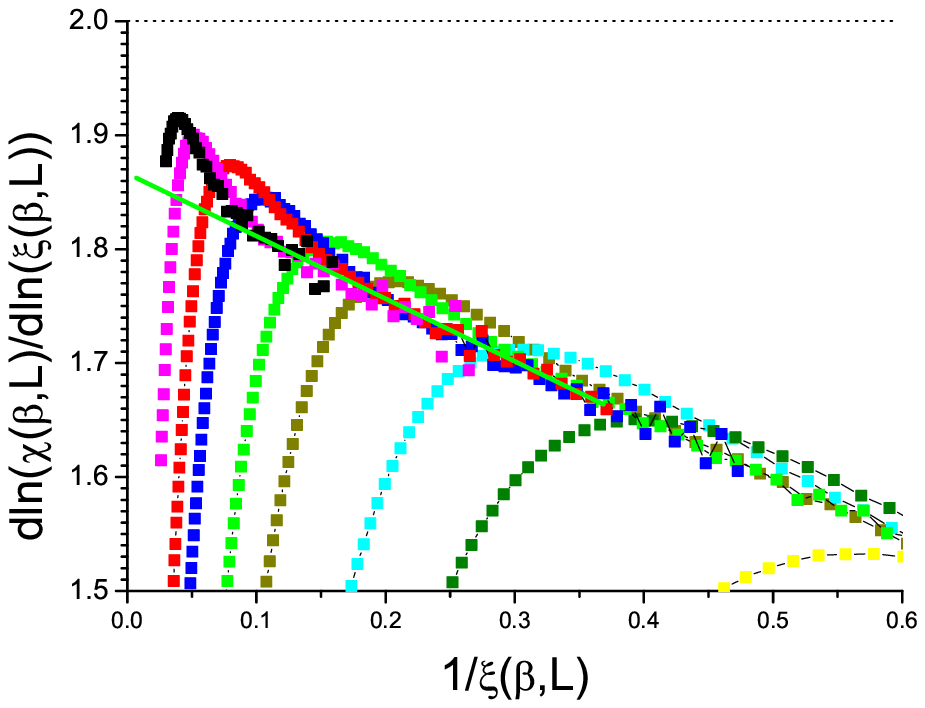}
  \caption{(Color on line) Symmetric Poisson 2D ISG. Logarithmic
    derivative $\partial\ln\chi(\beta,L)/\partial\ln\xi(\beta,L)$
    against $1/\xi(\beta,L)$ for $L= 128$, $96$, $48$, $32$, $24$,
    $16$, $12$, $8$ (left to right).} \protect\label{fig19}
\end{figure}

In Figs.~\ref{fig17}, \ref{fig18} and \ref{fig19} we show the
$y(\beta,L) = \partial\ln\chi(\beta,L)/\partial\ln\xi(\beta,L)$
against $x(\beta,L) = 1/\xi(\beta,L)$ plots for the diluted bimodal,
the "anti-diluted" bimodal and the symmetric Poisson model.  By mild
extrapolation the intercepts can be estimated to be $y(x=0) \approx
1.845$, $1.87$ and $1.90$ , i.e. $\eta = 2- y(x=0) \approx 0.155(10)$,
$0.13(1)$ and $0.10(1)$ for these models, weaker than the estimate
$\eta = 0.20(2)$ for the bimodal model \cite{lundow:16}, but still far
from zero. As in the bimodal ISG, there are overshoots as functions of
temperature for individual $L$ curves. (In Ref.~\cite{hartmann:08},
for a diluted bimodal model at zero temperature the estimate obtained
was $\eta \approx 0.20$).


\begin{figure}
  \includegraphics[width=3.5in]{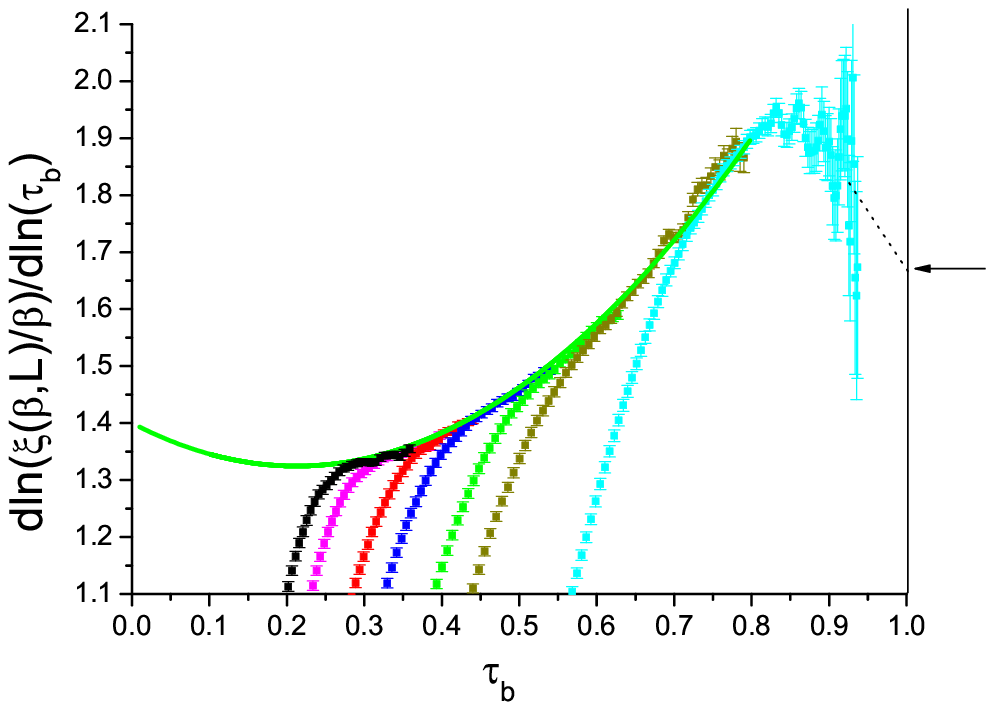}
  \caption{(Color on line) Diluted bimodal 2D ISG. Logarithmic
    derivative $\partial\ln[\xi(\beta,L)/\beta]/\partial\ln\tau_{b}$
    against $\tau_{b}$ for $L= 128$, $96$, $64$, $48$, $32$, $24$,
    $12$ (left to right). The continuous (green) curve is an
    extrapolated fit. The right hand side arrow indicates the exact
    infinite temperature limit.}  \protect\label{fig20}
\end{figure}

\begin{figure}
  \includegraphics[width=3.5in]{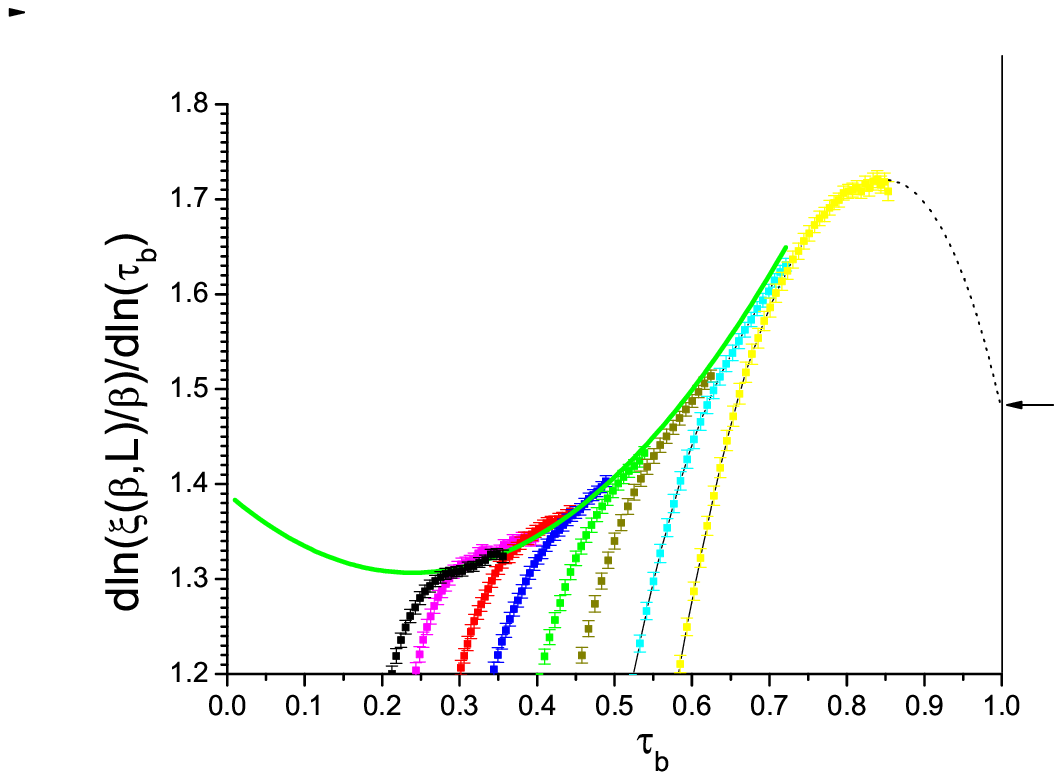}
  \caption{(Color on line) Anti-diluted bimodal 2D ISG. Logarithmic
    derivative $\partial\ln[\xi(\beta,L)/\beta]/\partial\ln\tau_{b}$
    against $\tau_{b}$ for $L= 128$, $96$, $48$, $32$, $24$, $16$,
    $12$ (left to right). The continuous (green) curve is an
    extrapolated fit. The right hand side arrow indicates the exact
    infinite temperature limit.}  \protect\label{fig21}
\end{figure}

\begin{figure}
  \includegraphics[width=3.5in]{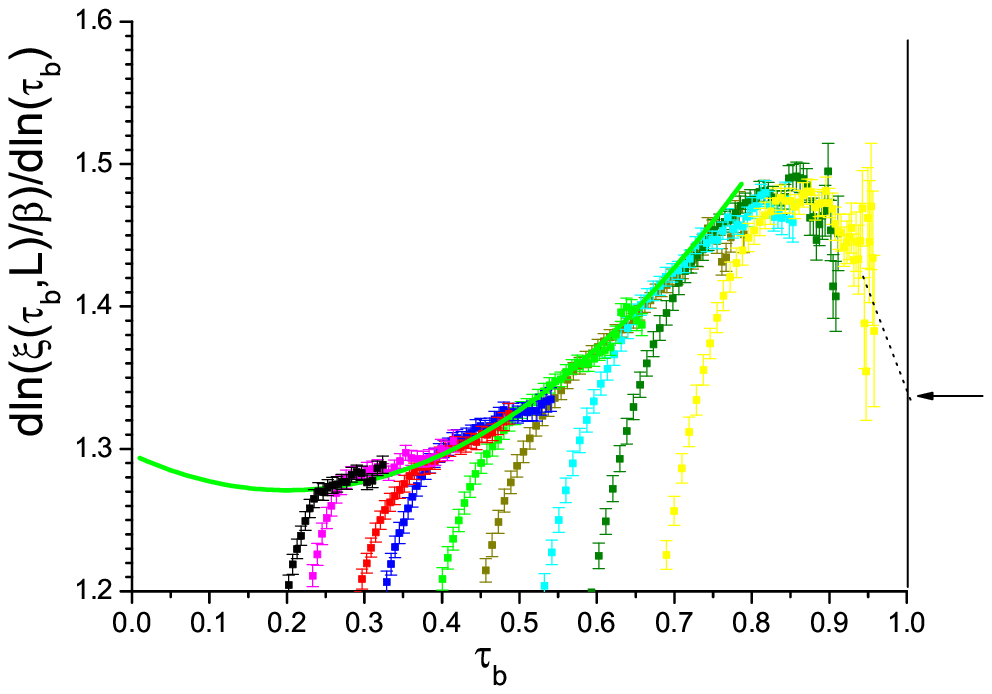}
  \caption{(Color on line) Symmetric Poisson 2D ISG. Logarithmic
    derivative $\partial\ln[\xi(\beta,L)/\beta]/\partial\ln\tau_{b}$
    against $\tau_{b}$ for $L= 128$, $96$, $48$, $32$, $24$, $16$,
    $12$, $8$ (left to right). The continuous (green) curve is an
    extrapolated fit. The right hand side arrow indicates the exact
    infinite temperature limit.}  \protect\label{fig22}
\end{figure}

\begin{figure}
  \includegraphics[width=3.5in]{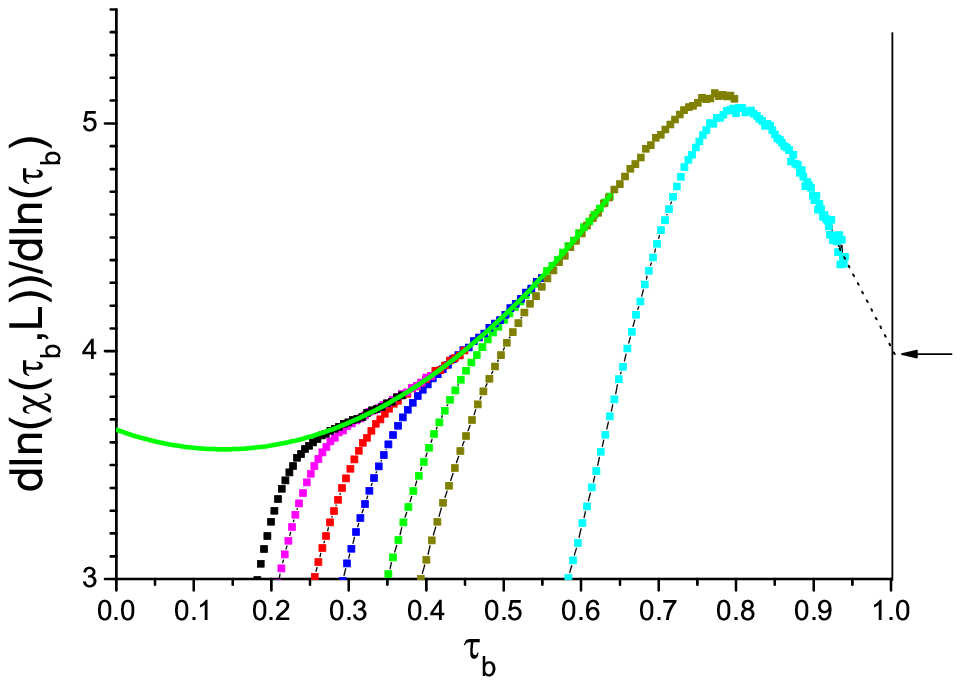}
  \caption{(Color on line) Diluted bimodal 2D ISG. Logarithmic
    derivative $\partial\ln\chi(\beta,L)/\partial\ln\tau_{b}$ against
    $\tau_{b}$ for $L= 128$, $96$, $48$, $32$, $24$ (left to
    right). The continuous (green) curve is an extrapolated fit. The
    right hand side arrow indicates the exact infinite temperature
    limit.}  \protect\label{fig23}
\end{figure}

\begin{figure}
  \includegraphics[width=3.5in]{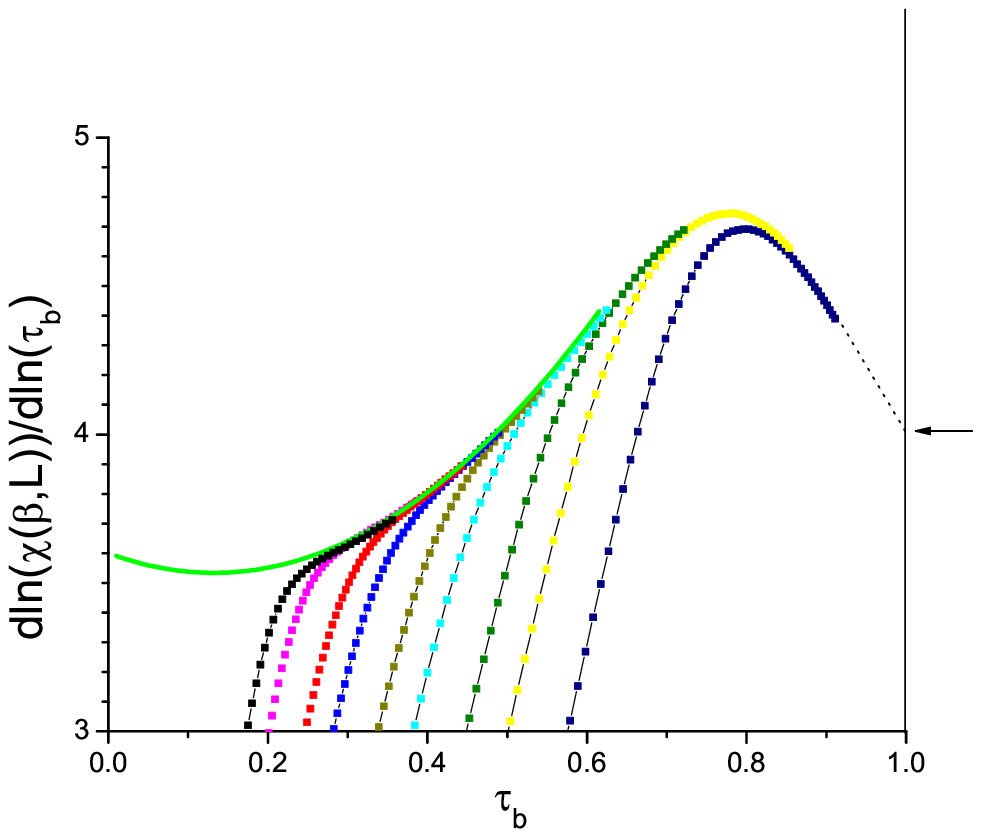}
  \caption{(Color on line) Anti-diluted bimodal 2d ISG. Logarithmic
    derivative $\partial\ln\chi(\beta,L)/\partial\ln\tau_{b}$ against
    $\tau_{b}$ for $L= 128$, $96$, $48$, $32$, $24$, $16$, $12$ (left
    to right). The continuous (green) curve is an extrapolated
    fit. The right hand side arrow indicates the exact infinite
    temperature limit.}  \protect\label{fig24}
\end{figure}

\begin{figure}
  \includegraphics[width=3.5in]{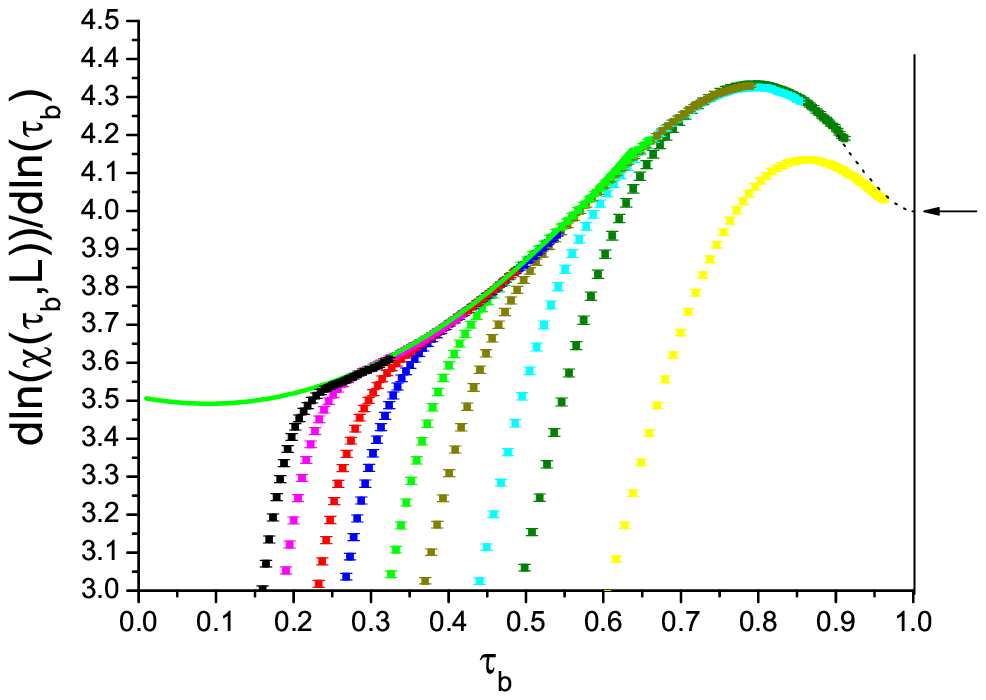}
  \caption{(Color on line) Symmetric Poisson 2d ISG. Logarithmic
    derivative $\partial\ln\chi(\beta,L)/\partial\ln\tau_{b}$ against
    $\tau_{b}$ for $L= 128$, $96$, $48$, $32$, $24$, $16$, $12$, $8$
    (left to right). The continuous (green) curve is an extrapolated
    fit. The right hand side arrow indicates the exact infinite
    temperature limit.}  \protect\label{fig25}
\end{figure}

In Figs.~\ref{fig20}, \ref{fig21}, and \ref{fig22} we show the
effective exponent $\nu_{b}(\beta,L) =
\partial\ln[\xi(\beta,L)/\beta]/\partial\ln\tau_{b}$ for all sizes $L$
for these models, and in Figs.~\ref{fig23}, \ref{fig24} and
\ref{fig25} we show the effective exponents $\gamma_{b}(\beta,L) =
\partial\ln\chi(\beta,L)/\partial\ln\tau_{b}$.  We have carried out
extrapolations using just the same polynomial fit procedure as
explained in \cite{lundow:16} and in the Appendix in order to estimate
the zero temperature critical intercepts.  

The extrapolated critical exponent estimates for the diluted bimodal
model, the anti-diluted bimodal model, and the symmetric Poisson model
are $\nu_{b} = 1.40(2)$, $1.39(2)$, $1.30(2)$ and $\gamma_{b} =
3.65(5)$, $3.60(5)$, $3.46(2)$ respectively, as compared with $\nu_{b}
= 1.9(1)$, $\gamma_{b} = 4.3(1)$ for the bimodal model
\cite{lundow:16}.  These exponents are related to the correlation
length critical exponent $\nu$ in the traditional $T$ scaling
convention by $\nu_{b} = (\nu - 1)/2$ and $\gamma_{b} = \nu(2-\eta)/2$
\cite{lundow:16}. Thus the critical exponent estimates for the
degenerate ground state models are consistent with $\eta = 0.155(5)$,
$\nu = 3.8(1)$, $\eta = 0.13(1)$, $\nu = 3.7(2)$, and $\eta =
0.10(2)$, $\nu = 3.6(1)$ respectively, as compared with
$\eta=0.20(2)$, $\nu = 4.8(1)$ for the bimodal model (and $\eta = 0$,
$\nu = 3.55(2)$ for the continuous distribution models).  The data for
the bimodal model true correlation length at low temperatures obtained
by Merz and Chalker with a remarkable network mapping technique,
Ref.~\cite{merz:02} Fig.~24, can be extrapolated to a critical
exponent value $\nu \approx 4.6$ which is consistent with the
simulation estimate for the bimodal value $\nu$ in
Ref.~\cite{lundow:16}.


Although these values are similar to each other they are
all different and all are quite distinct from the bimodal model
estimates $\eta = 0.20(2), \nu = 4.8(3)$, \cite{lundow:16}.

\section{Conclusions}\label{sec:VII}
We show simulation data for three continuous and four discrete
interaction distribution 2D ISG models and for the 2D fully frustrated
Villain model (Appendix I).  All these models order only at zero
temperature. The simulation techniques and the analysis follow
strictly those of Ref.~\cite{lundow:16} where results for the
canonical 2D ISG bimodal (discrete) and Gaussian (continuous)
interaction distribution models were reported. We have made extensive
simulation measurements up to size $L=128$ on each model, which have
been analysed using the 2D scaling parameter $\tau_{b} =
1/(1+\beta^2)$ as in \cite{lundow:16} as well as the traditional
scaling parameter $T$.

In the class of ISG models with continuous interaction distributions,
in addition to the Gaussian distribution we have studied the uniform
interaction distribution and the Laplacian interaction
distribution. These models have non-degenerate ground states and as a
consequence an anomalous dimension exponent $\eta \equiv 0$. Except
for very small sizes and high temperatures, for all $\eta=0$ models
and for all $L$, Binder parameter $U_{4}(\beta,L)$ against normalized
second moment correlation length $\xi(\beta,L)/L$ data lie on a single
universal curve extending to the zero temperature limit $[U_{4}(L)=1,
  \xi(L)/L \equiv \infty]$.

The present numerical data show that estimates for the critical second
moment correlation length exponent for the continuous interaction
distributions are all compatible with $\nu = 3.55(2)$ (expressed in
terms of the $T$ temperature scaling convention), which is the
accepted value for the Gaussian distribution 2D ISG
\cite{rieger:97,hartmann:02,carter:02,hartmann:02a,houdayer:04}. This
result is consistent with all continuous interaction distribution 2D
ISGs forming a single universality class.

The bimodal interaction 2D ISG, a diluted bimodal interaction 2D ISG,
an "anti-diluted" 2D ISG, a multi-peak 2D ISG, and the 2D FF model,
all order only at zero temperature, have discrete interaction
distributions, and have highly degenerate ground states. For each
model the specific heat data show crossovers at size dependent
temperatures $T^{*}(L)$ between an effectively continuous energy state
distribution regime for $T > T^{*}(L)$ and a ground state plus excited
state dominated regime for $T < T^{*}(L)$ .  For each of these models,
Binder parameter $U_{4}(\beta,L)$ against normalized second moment
correlation length $\xi(\beta,L)/L$ data do not lie on the $\eta=0$
universal curve, and for every $L$ the data tend to zero temperature
end-points which are far from $U_{4}(L)=1$, $\xi(L)/L \equiv
\infty$. As the temperature is lowered the data evolve continuously
and smoothly through $T^{*}$ indicating that the effective $\eta$
values in the $T > T^{*}(L)$ and $T < T^{*}(L)$ regimes are the
same. The end point values of $\xi(L)/L$ extrapolated to infinite $L$
are different for each model, implying that the models all lie in
different universality classes with different non-zero $\eta$ values.

From scaling analyses, the critical exponents of the discrete
distribution ISGs are estimated to be $\eta=0.20(2)$, $\nu=4.8(1)$ for
the bimodal model, $\eta=0.155(5)$, $\nu=3.8(1)$ for the $p=0.125$
diluted bimodal model, $\eta=0.13(1)$, $\nu=3.8(1)$ for the $p=0.20$
anti-diluted bimodal model, and $\eta=0.10(1)$, $\nu=3.6(1)$ for the
symmetric Poisson model defined above.

Each of the present discrete distribution models represents an
infinite family of possible models. If a parameter defining a
particular model was modified (for instance by choosing other values
of $p$ for the diluted or anti-diluted models) we would expect the
critical exponents to change continuously as functions of $p$,
starting of course from the bimodal values for $p=0$.


To summarize, the 2D ISG models with continuous interaction
distributions lie in a single universality class, but the 2D ISG
models with discrete distributions do not share this universality
class. On the contrary each discrete distribution model has its
individual critical exponents.


When it was reported in 1980 by Morgenstern and Binder
\cite{morgenstern:80} that the 2D bimodal ISG had a value $\eta =
0.4(1)$ which is non-zero so different from the $\eta \equiv 0$ of the
Gaussian model, it was suggested that this universality breakdown
behavior could arise from higher order terms in the
$\epsilon$-expansion for the critical exponents in dimensions below
upper critical dimension $d=6$ \cite{bray:81}, see also
\cite{elderfield:78}. Indeed there is now numerical evidence for
non-universality in dimensions $d=4$ \cite{lundow:15,lundow:15a} and
$d=5$ \cite{lundow:16a} as well as in dimension $d=2$.

\begin{acknowledgments}
  We thank Helmut Katzgraber for generously giving us access to all
  the raw Fully Frustrated data originally generated for
  Ref.~\cite{katzgraber:08}, and to the raw low temperature bimodal
  ISG data originally generated for Ref.~\cite{katzgraber:07}. We
  would like to thank Mike Moore for pointing out references
  \cite{bray:81}, \cite{elderfield:78} and \cite{merz:02}, and John
  Chalker for helpful comments.  The computations were performed on
  resources provided by the Swedish National Infrastructure for
  Computing (SNIC) at the Chalmers Centre for Computational Science
  and Engineering (C3SE).
\end{acknowledgments}

\appendix

\section{The fully frustrated Villain model}\label{sec:A}
In the square lattice fully frustrated (FF) Villain model
\cite{villain:77} all near neighbor interactions have strength $|J|$;
in the $x$ direction all bonds are ferromagnetic, while in the $y$
direction columns of bonds are alternately ferromagnetic and
antiferromagnetic, so every plaquette is frustrated. This is a well
understood 2D model with a zero temperature ferromagnetic transition
and a strong ground state degeneracy, which can provide a basis of
comparison for other models with ground state degeneracies such as
discrete interaction distribution ISG models.

\begin{figure}
  \includegraphics[width=3.5in]{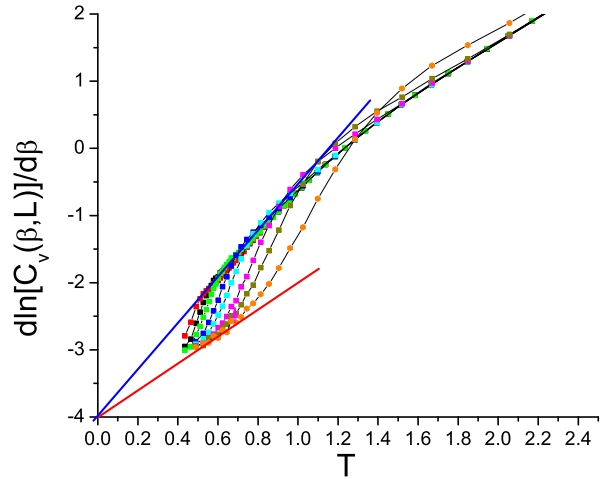}
  \caption{(Color on line) Fully Frustrated 2D model. The logarithmic
    derivative of the specific heat
    $\partial\ln C_{v}(\beta,L)/\partial\ln\beta$ against temperature
    $T$. $L= 96$, $64$, $48$, $32$, $24$, $16$, $12$, $8$ (left to
    right). Upper blue straight line : the thermodynamic limit $T >
    T^{*}(L)$ envelope curve $y(x) = -4 +3.5x$. Lower red straight
    line : the finite size ground state dominated $T < T^{*}(L)$
    regime $y(x) = -4 +2x$. } \protect\label{fig26}
\end{figure}

For the FF model a number of properties have been established
analytically \cite{forgacs:80}, by precise energy measurements
\cite{lukic:06}, and by simulations \cite{katzgraber:08}. The FF
ground state degeneracy corresponds to a zero temperature entropy per
site of $0.2916$ \cite{forgacs:80}. (For comparison in the 2D bimodal
ISG the zero temperature entropy per site is $0.078(5)$
\cite{hartmann:02,poulter:05,thomas:07}). The first FF excited states
are at $4J$. The zero temperature FF ordering is ferromagnetic, with a
thermodynamic limit ($L=\infty$, $T=0$) anomalous dimension exponent
$\eta \equiv 1/2$ \cite{forgacs:80} and a low temperature
thermodynamic limit second moment correlation length $\xi(\beta) \sim
\exp(2\beta)/2$ \cite{forgacs:80,lukic:06,katzgraber:08}.  The FF
specific heats in the infinite $L$ and finite $L$ limits were
estimated in Ref.~\cite{lukic:06} by sophisticated Pfaffian algebra to
be of the form $C_{v} \sim \beta^{B}\exp(-A\beta)$, the values being
$B=3$ in the infinite $L$ limit and $B=2$ in the finite $L$ limit with
$A=4$ in both limits. We show in Fig.~\ref{fig26} FF specific heat
data for a wide range of sizes in the form $y(\beta,L) = \partial\ln
C_{v}(\beta,L)/\partial\beta$ against $x = T$. This type of plot leads
to a straight line with intercept $-A$ and slope $B$. For finite sizes
in the FF model there is a crossover at a size dependent temperature
$T^{*}(L)$, just as in the 2D bimodal ISG \cite{jorg:06,thomas:11}.
The FF $T^{*}(L)$ crossover from an effectively continuous energy
level regime to the ground state plus gap dominated regime can be
identified by inspection of Fig.~\ref{fig26} as the region where for
each $L$ the curve $y(x)$ passes from the thermodynamic limit $T >
T^{*}(L)$ envelope curve $y(x) \approx -4 +3.5x$ to the finite size
ground state dominated regime $T < T^{*}(L)$ line $y(x) = -4 +2x$. The
$A$ and $B$ values practically agree with Ref.~\cite{lukic:06}; the
crossover temperatures are near $T^{*}(L) \approx 0.5$. The present
figure can be compared directly to the equivalent figure for the 2D
bimodal ISG, Ref.~\cite{lundow:16} Fig.~2. The lower diagonal line in
the present Fig.~\ref{fig26} corresponds to just the same "na\"\i ve"
ground state plus $4J$ gap dominated specific heat regime as in the 2D
bimodal ISG, $C_{v}(T) \sim \exp(-4/T)/T^2$, but the 2D bimodal ISG
large $L$ thermodynamic limit specific heat curve with $A \approx 0$
and $B$ negative is very different from the FF large $L$ limit curve.

\begin{figure}
  \includegraphics[width=3.5in]{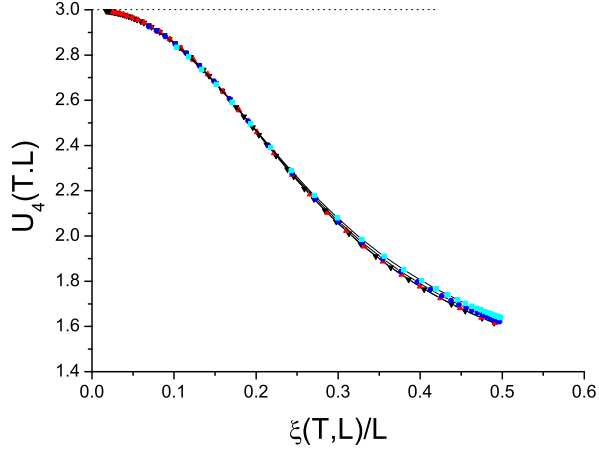}
  \caption{(Color on line) Fully Frustrated 2D model. The low
    temperature Binder cumulant $U_{4}(T,L)$ against the normalized
    correlation length $\xi(T,L)/L$. $L=8$ (cyan squares), $L=12$
    (blue circles), $L= 32$ (red triangles), $L=48$ (black inverted
    triangles). For each $L$ the data points terminate smoothly at a
    zero temperature end-point.}  \protect\label{fig27}
\end{figure}

The FF $U_4(\beta,L)$ against $\xi(\beta,L)/L$ curve breaks off rapidly
from the $\eta=0$ universal curve to arrive smoothly at a critical
end-point $\xi(T=0,L)/L = 0.488 +0.1/L$, $U_4(T=0,L)=1.618-0.2/L)$
\cite{katzgraber:08}, Fig.~\ref{fig27}.

\begin{figure}
  \includegraphics[width=3.5in]{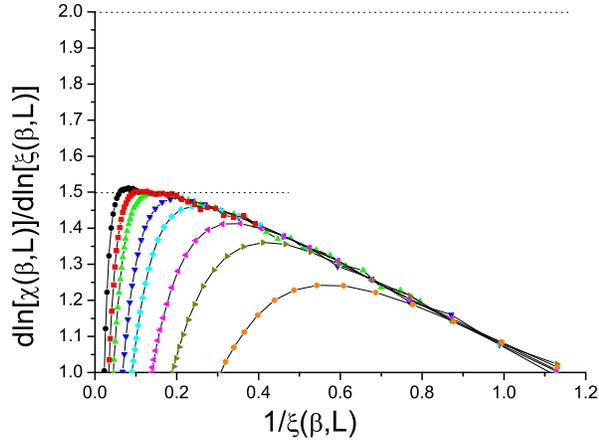}
  \caption{(Color on line) Fully Frustrated 2D model. The logarithmic
    derivative $\partial\ln\chi(\beta,L)/\partial\ln\xi(\beta,L)$
    against $1/\xi(\beta,L)$. $L= 96$, $64$, $48$, $32$, $24$, $16$,
    $12$, $8$ (left to right).}  \protect\label{fig28}
\end{figure}

In Fig.~\ref{fig28}, we show the FF derivative $y(\beta,L) =
\partial\ln\chi(\beta,L)/\partial\ln\xi(\beta,L)$ against $x(\beta,L)
= 1/\xi(\beta,L)$, where $\xi(\beta,L)$ is the second moment
correlation length and $\chi(\beta,L)$ is the susceptibility. In the
present Fig.~\ref{fig28} (as in the bimodal and Gaussian ISG figures
in Ref.~\cite{lundow:16}, Figs.~3 and 4) for all the ThL envelope
points the data are in the regime $T > T^{*}(L)$.

The $L$-independent envelope curve of all the FF data in the ThL
regime $L > \xi(\beta,L)$, $T > T^{*}(L)$ can be identified by
inspection. The essential point is that the "high temperature" regime
FF ThL derivative $y(\beta,L)$ from temperatures above the crossovers
extrapolates smoothly and accurately to $y(\beta,L) = 1.5$, so to
$y(\beta,L) = 2-\eta$ with an effective limiting $\eta$ equal to
$1/2$, the analytically known $L=\infty$, $T = 0$ critical exponent
\cite{forgacs:80}.


Thus in the FF model, it is found that when the "effectively
continuous energy level" regime effective exponent $\eta(T,L)$ is
extrapolated to the limit of large $L$ using the ThL
$\partial\ln\chi(\beta,L)/\partial\ln\xi(\beta,L)$ differentiation
procedure, the value is equal to the $T \equiv 0$ ground state
critical exponent.  This can be taken to imply that there is no
difference between these two limiting exponent values in the discrete
interaction distribution ISG models either.

\section{Fitting procedure}\label{sec:B}

In Ref.~\cite{lundow:16} the data for the derivative of the
susceptibility and the second moment correlation length
$\gamma_{b}(\tau_{b},L) =
\partial\ln\chi(\beta,L)/\partial\ln\tau_{b}$ and
$\nu_{b}(\tau_{b},L) =
\partial\ln[\xi(\beta,L)/\beta]/\partial\ln\tau_{b}$ were extrapolated
to $\tau_{b}=0$ after making three parameter polynomial fits of the
type $y(\tau_{b}) = a+b\tau_{b}+c\tau_{b}^2$.

\begin{figure}
  \includegraphics[width=3.5in]{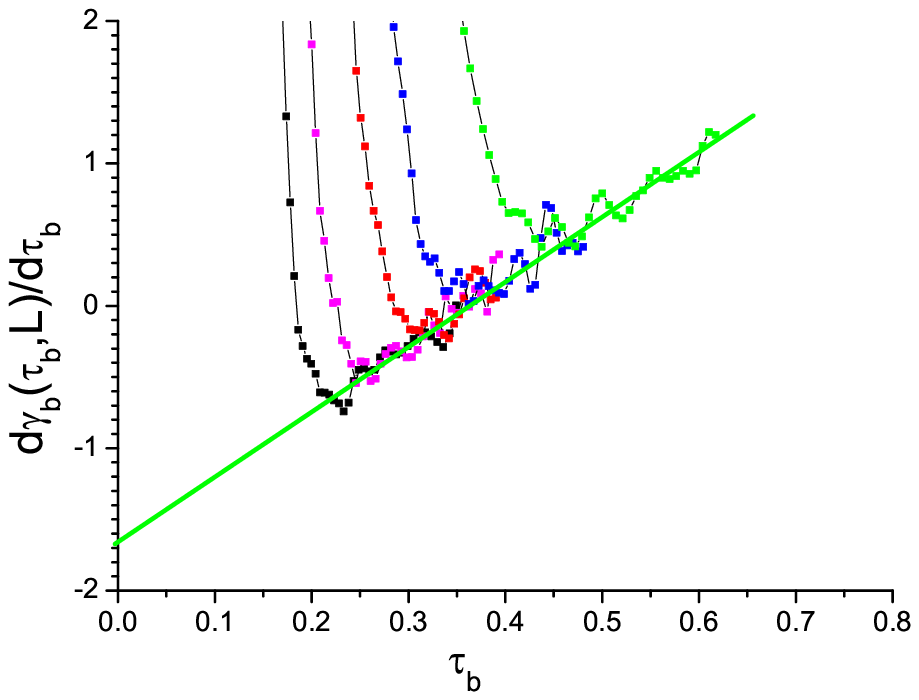}
  \caption{(Color on line) Gaussian 2D ISG. The derivative
    $\partial\gamma_{b}(\tau_{b},L)/\partial\tau_{b}$ against
    $\tau_{b}$ for $L= 128$, $96$, $64$, $48$, $32$ (left to right).
    Straight green line : fit to the ThL regime data}
  \protect\label{fig29}
\end{figure}

\begin{figure}
  \includegraphics[width=3.5in]{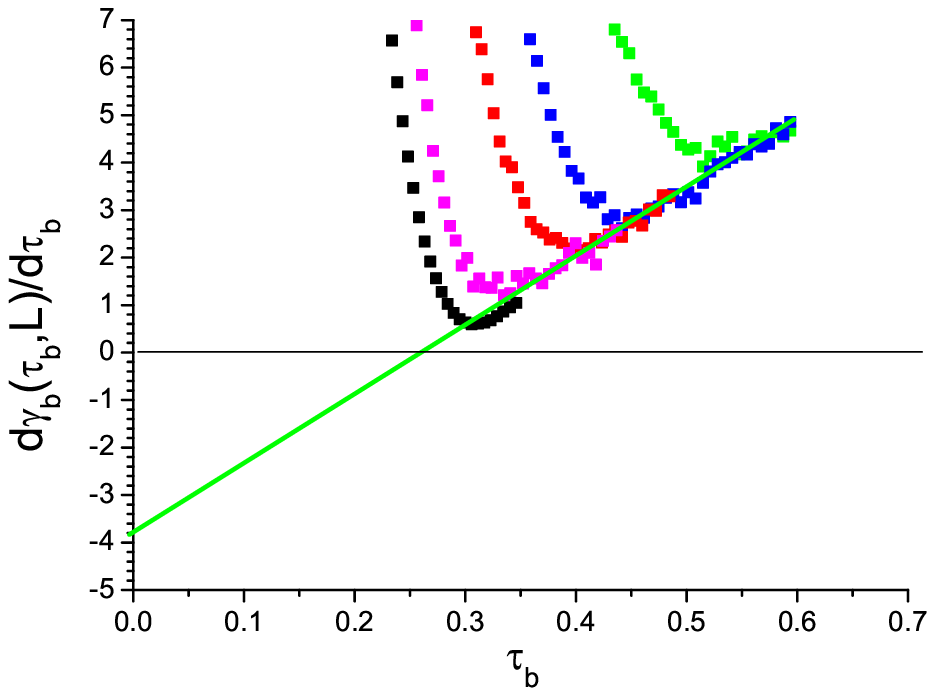}
  \caption{(Color on line) Bimodal 2D ISG. The derivative
    $\partial\gamma_{b}(\tau_{b},L)/\partial\tau_{b}$ against
    $\tau_{b}$ for $L= 128$, $96$, $64$, $48$, $32$ (left to right).
    Straight green line : fit to the ThL regime data}
  \protect\label{fig30}
\end{figure}

\begin{figure}
  \includegraphics[width=3.5in]{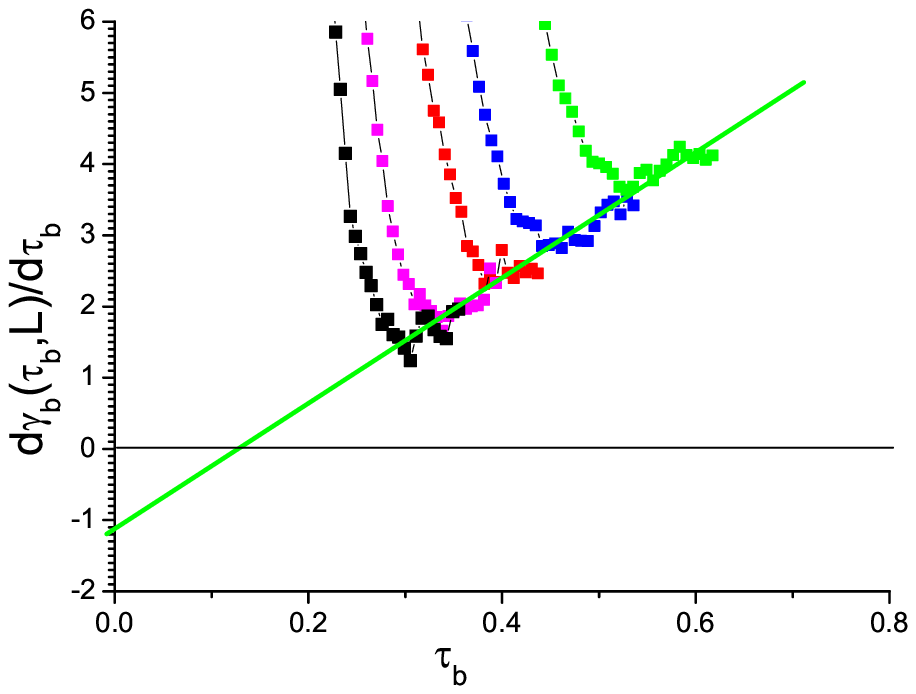}
  \caption{(Color on line) Diluted bimodal 2D ISG. The derivative
    $\partial\gamma_{b}(\tau_{b},L)/\partial\tau_{b}$ against
    $\tau_{b}$ for $L= 128$, $96$, $64$, $48$, $32$ (left to right).
    Straight green line : fit to the ThL regime data}
  \protect\label{fig31}
\end{figure}

\begin{figure}
  \includegraphics[width=3.5in]{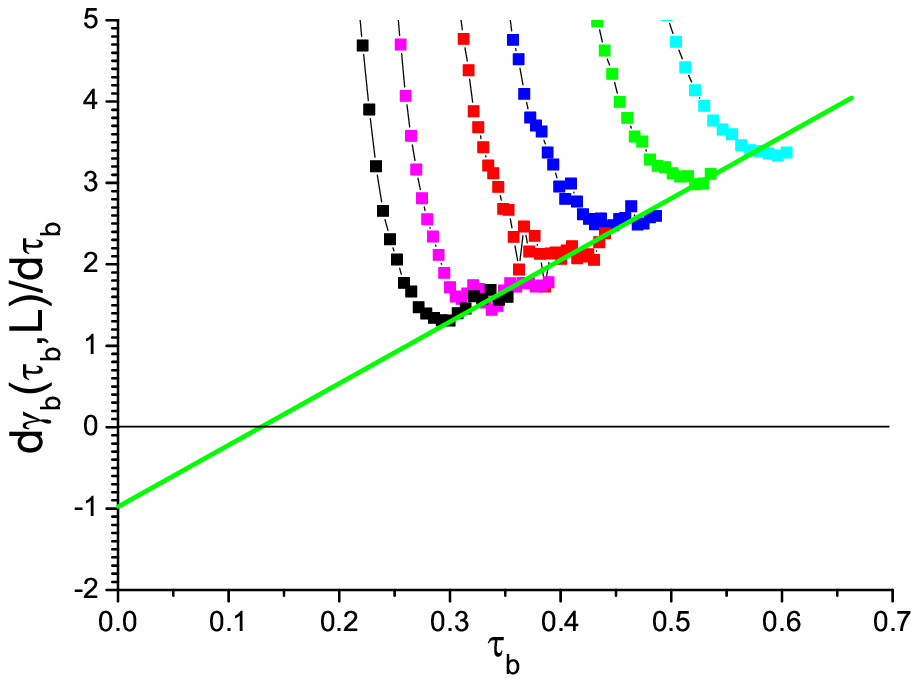}
  \caption{(Color on line) Anti-diluted bimodal 2D ISG. The derivative
    $\partial\gamma_{b}(\tau_{b},L)/\partial\tau_{b}$ against
    $\tau_{b}$ for $L= 128$, $96$, $64$, $48$, $32$, $24$ (left to
    right).  Straight green line : fit to the ThL regime data}
  \protect\label{fig32}
\end{figure}

\begin{figure}
  \includegraphics[width=3.5in]{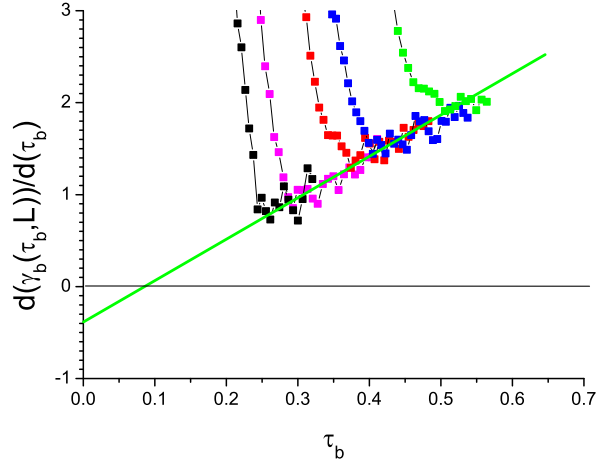}
  \caption{(Color on line) Symmetric Poisson 2D ISG. The derivative
    $\partial\gamma_{b}(\tau_{b},L)/\partial\tau_{b}$ against
    $\tau_{b}$ for $L= 128$, $96$, $64$, $48$, $32$ (left to right).  Straight
    green line : fit to the ThL regime data} \protect\label{fig33}
\end{figure}

In the present work we carry out the same type of fit but in two
stages.  First we plot the higher derivatives
$\partial\gamma_{b}(\tau_{b},L)/\partial\tau_{b}$ and
$\partial\nu_{b}(\tau_{b},L)/\partial\tau_{b}$ against $\tau_{b}$. In
each case a two parameter straight line fit $y(\tau_{b}) = b
+2c\tau_{b}$ to the ThL data up to about $\tau_{b} = 0.50$ is quite
acceptable. This implies that the leading Wegner correction exponent
$\theta$ happens to be close to $1.0$ in all models, as was assumed in
Ref.~\cite{lundow:16}, and justifies the simple polynomial fit
procedure.  Susceptibility
$\partial\gamma_{b}(\tau_{b},L)/\partial\tau_{b}$ data are shown in
Figs.~\ref{fig29}, \ref{fig30}, \ref{fig31}, \ref{fig32}, and
\ref{fig33}. The $\partial\nu_{b}(\tau_{b},L)/\partial\tau_{b}$ data
have a similar aspect but are intrinsically more noisy. With the
parameters $b$ and $c$ in hand for each model and so with a single
remaining free parameter, $a$, fits were made up to $\tau_{b} \approx
0.50$ to each of the $\gamma_{b}(\tau_{b},L)$ and
$\nu_{b}(\tau_{b},L)$ ThL curves shown in the earlier sections.

\end{document}